# A hybrid discrete-continuum approach to model hydro-mechanical behaviour of soil during desiccation

Khoa M. Tran[a], Ha H. Bui[a,*] and Giang D. Nguyen[b]

[a]Department of Civil Engineering, Monash University, Australia

[b]School of Civil, Environmental & Mining Engineering, The University of Adelaide, Australia

**Abstract**

Desiccation cracking in clayey soils occurs when they lose moisture, leading to an increase in their compressibility and hydraulic conductivity and hence significant reduction of soil strength. The prediction of desiccation cracking in soils is challenging due to the lack of insights into the complex coupled hydro-mechanical process at the grain scale. In this paper, a new hybrid discrete-continuum numerical framework, capable of capturing hydro-mechanical behaviour of soil at both grain and macro scales, is proposed for predicting desiccation cracking in clayey soil. In this framework, a soil layer is represented by an assembly of DEM particles, each occupies an equivalent continuum space and carries physical properties governing unsaturated flow. These particles move freely in the computational space following the discrete element method (DEM), while their contact network and the continuum mixture theory are used to model the unsaturated flow. The dependence of particle-to-particle contact behaviour on water content is represented by a cohesive-frictional contact model, whose material properties are governed by the water content. In parallel with the theoretical development is a series of experiments on 3D soil desiccation cracking to determine essential properties and provide data for the validation of mechanical and physical behaviour. Very good agreement in both physical behaviour (e.g. evolution of water content) and mechanical behaviour (e.g. occurrence and development of cracks, and distribution of compressive and tensile strains) demonstrates that the proposed framework is capable of capturing the hydro-mechanical behaviour of soil during desiccation. The capability of the proposed framework facilitates numerical experiments for insights into the hydro-mechanical behaviour of unsaturated soils that have not been possible before.

*Keywords*: Hydro-mechanical behaviour; Desiccation cracking; Crack pattern; DEM; Shrinkage

## 1. Introduction

Soil desiccation cracking often occurs in clayey soils when they lose moisture. When this process takes place, the soil strength decreases significantly, while the compressibility and hydraulic conductivity increase. These changes in the soil properties significantly affect the stability of adjacent structures and the performance of geoengineering applications (Morris et al. 1992, Teodosio et al. 2020, Tran et al. 2020). Therefore, understanding soil desiccation cracking is very important.

*Corresponding author
Email addresses: ha.bui@monash.edu (Ha Bui), khoa.tran1@monash.edu (Khoa Tran), g.nguyen@adelaide.edu.au (Giang Nguyen). This manuscript was accepted to published in *Journal of Geotechnical and Geoenvironmental Engineering (ASCE)*.

Experimental studies have reported that the desiccation cracking process of clayey soils generally takes place in three stages: the primary, secondary and steady stages. In the primary stage, several cracks initiate and propagate in the soil until they reach each other. In the secondary stage, more cracks appear, starting from pre-existing cracks and terminating after reaching other pre-existing cracks, dividing the soil media into smaller fragments. In the steady stage, only a few new cracks occur, while pre-existing cracks widen (Corte and Higashi 1960, Ghazizade and Safari 2017, Konrad and Ayad 1997, Nahlawi and Kodikara 2006, Shin and Santamarina 2011, Tang et al. 2011). However, during desiccation, clayey soil shrinks significantly, sensors embedded in soil samples lose their contact with soil (Cheng et al. 2020), and thus the hydro-mechanical behaviour of these samples (e.g. changes in stress, strain and water content inside samples) cannot be fully captured and understood. Therefore, all explanations of the initiation and development of cracks based on experimental results are qualitative. For example, by looking at the occurrence of cracks in samples during drying under different boundary conditions, cracks are explained to occur because the tensile stress reaches the tensile strength of the soil. This tensile stress is assumed to develop in these samples due to movements of soil particles and soil layers are hindered because of friction and adhesion, which exist between the soil and external boundaries or lower soil layers (Corte and Higashi 1960, Kodikara and Costa 2013, Morris et al. 1992, Peron et al. 2009). However, by looking at the direction of the horizontal strain on the sample surface, Wei et al. (2016) concluded that cracks could occur due to tensile failure or a combination of both tensile and shear failures. Another example is the explanation based on the intrusion process of the air-water interface. This example was proposed mainly based on the calculation of the degree of saturation of these samples during drying (Shin and Santamarina 2011, Tang et al. 2011). Therefore, further studies are required to provide further insights into the hydro-mechanical behaviour of soil during desiccation.

Numerical methods have been employed to investigate soil desiccation cracking. In the context of continuum numerical approach, Trabelsi et al. (2012) implemented a simple elasto-visco-plasticity model in a finite element program to simulate cracking patterns in soil samples during experiments. Although cracks occurred in numerical samples, the width of the cracks did not change during desiccation, and some cracks did not connect to form clear crack patterns. Peron et al. (2013) used the total stress analysis framework based on Biot theory and finite element method (FEM) to explain the initiation and development of cracks. However, the development of cracks could not be handled in their work due to the limitation of their FEM model. This framework was further developed by Vo et al. (2017) to capture hydro-mechanical behaviour of soils as well as the occurrence of cracks during drying by using a damage-plasticity model and interface elements (or joint elements). The results showed that the framework could fairly capture the hydro-mechanical behaviour of soils and the occurrence of cracks during soil desiccation. However, only the two-dimensional vertical crack network occurred through prescribed vertical interface elements, while complex networks that often occurred in nature were not developed. To overcome this problem, a technique known as "mesh fragmentation" introduced by Sánchez et al. (2014) were used in some recent studies. Their results showed that most of the cracking behaviour and complex crack networks observed in laboratory and field tests during desiccation could be captured using this technique (Maedo et al. 2020, Sánchez et al. 2014). Similarly, other continuum numerical methods such as the particle discretisation scheme finite element method (PDS-FEM), the finite different method (FDM) and the smoothed particle hydrodynamics method (SPH) were also employed to investigate the occurrence of cracks during soil desiccation (Bui et al. 2015; Hirobe and Oguni 2016, Stirling et al. 2017, Tran et al. 2019; Tran et al. 2020). The results showed that 2D/3D crack patterns were fairly well reproduced. However, due to lacking the behaviour at meso and grain scales, all continuum methods cannot explain all phenomena observed in clayey soil during desiccation, such as the development and distribution of both compressive and tensile strains.

In contrast to the continuum approach, the discrete element method, which has been widely applied to study macro- and micro-behaviours of soils such as behaviour of soils due to mass loss (Lee et al. 2012, Tran et al. 2011, Tran et al. 2012), behaviour of unsaturated soils (Anandarajah and Amarasinghe 2012, Tran et al. 2018) and behaviour of granular soils (Bui et al. 2009, Jarrar et al. 2020, Nguyen et al. 2020), was also employed to study soil cracking (Guo et al. 2017, Peron et al. 2009, Sima et al. 2014). In these studies, bonded DEM particles were used to represent clay aggregates. The drying process was implemented by changing the size of the DEM particles based on the water content and volume strain of experimental samples. Other model parameters (i.e. stiffness, tensile strength and shear strength for soil–soil and soil–mould interactions) were calibrated from experimental results obtained in uniaxial tensile tests or uniaxial compression tests. Results of these studies demonstrated that the DEM was a potential approach for studying soil desiccation cracking, as cracks can occur naturally in DEM and multiple cracking networks developed during desiccation could be well captured by DEM. However, owing to the simple contact bond model adopted in these studies, mixed-mode failure (the combination of tensile failure and shear failure) observed in clay soils during the desiccation process (TRAN 2019, Wei et al. 2016) could not be captured. Furthermore, none of the existing DEM studies has been capable of capturing the hydro-mechanical process of soil desiccation cracking and thus the shrinkage process of soil layers (e.g. the soil layers detach from the mould and then macro cracks start occurring) does not naturally occur. Therefore, further study is required to account for the hydro-mechanical behaviour of soils and to properly simulate the shrinkage process as well as to capture the mixed-mode failure when soil desiccation cracking is analysed using the DEM.

In this paper, a hydro-mechanical DEM framework is developed to study soil desiccation cracking. The unique feature of this framework is its capability to simulate unsaturated flow at all different saturation regimes through deformed porous media, as well as to capture the shrinkage process and mixed-mode failure of cracks in soil media during desiccation. This paper begins with the development of the hydro-mechanical framework, followed by a series of shrinkage experiments to provide data for the validation process of the framework, the performance of the framework and insights into the behaviour of clayey soils during drying.

## 2. The hydro-mechanical DEM framework

In this section, general assumptions are first given to provide the theoretical basis for deriving the mathematical equations of the proposed hydro-mechanical DEM framework. The governing equations for both hydraulic and mechanical behaviour of soils are then presented. Finally, constitutive relations and numerical algorithms for coupling the hydraulic behaviour and mechanical behaviour are established.

## 2.1. General assumptions

Previous studies have reported that a clay soil layer is an assembly of clay aggregates, which are formed by several clay particles tied together due to inter-particle forces (e.g. van der Waal force, double-layer force) (Mitchell and Soga 2005, Wang and Xu 2007) (Fig. 1(a)). During desiccation, the water content of the soil layer decreases and this layer shrinks due to the rearrangement of clay particles or shrinkage of clay aggregates (Sima et al. 2014), while the suction and tensile strength of the soil layer increase. Tensile stresses within the soil layer also develop as the movement of soil particles is hindered by friction and adhesion at the bottom of the soil layer. Consequently, cracks occur when the tensile stresses inside the soil specimen reach the tensile strength of the soil (Corte and Higashi 1960, Kodikara and Costa 2013, Morris et al. 1992, Peron et al. 2009). To model the mechanical behaviour of clay soils during desiccation, the following assumptions are adopted:

- A clay layer is represented by an assembly of DEM particles, each of which represents a clay aggregate. These particles are bonded at their contacts via a system of normal and tangential springs (Fig. 1(b)). These springs can bear tensile, shear and compressive forces that represent the mechanical and hydraulic forces (e.g. friction, suction and bonding forces) in contact zones. These springs break when the stresses in the contact zones satisfy a yield criterion that considers changes in both normal and shear stresses. Moreover, the stiffness, tensile and shear strengths of these springs are governed by the water content of the soil layer.

- During drying, water evaporates and clay particles in each clay aggregate rearrange, leading to its volume reduction. This process is naturally simulated by reducing the size of DEM particles in the soil specimen. The rate of volumetric shrinkage of each DEM particle is proportional to the moisture evaporation rate occurred on each particle.

To model moisture evaporation in the deformable porous media, the discrete modelling numerical approach recently proposed by Tran et al. (2019) to simulate unsaturated seepage flows through non-deformable porous media is further extended to account for the influence of volumetric deformation on the seepage flow behaviour. The main assumptions of this approach are:

- Each DEM particle has its own solid volume, representing the solid volume of a clay aggregate, carries unsaturated flow information (e.g. hydraulic conductivity and volumetric water content) and occupies an equivalent continuum space, which can be approximated based on the solid volume and surrounding porosity (Fig. 1(c)).

- Water flows from the centre of the particle with higher volumetric water content (or higher piezometric head) to the centre of the particle with lower volumetric water content (or lower piezometric head) (Fig. 1(c)) through the contact point of these particles. This assumption represents an attempt to mimic the water flow process at a microscopic scale, in which water flows through water layers adhering to particles (Gili and Alonso 2002).

- Clay particles and water are assumed to be incompressible. Water is assumed to be Newtonian fluid, and no mass exchange between two phases are considered in this work.

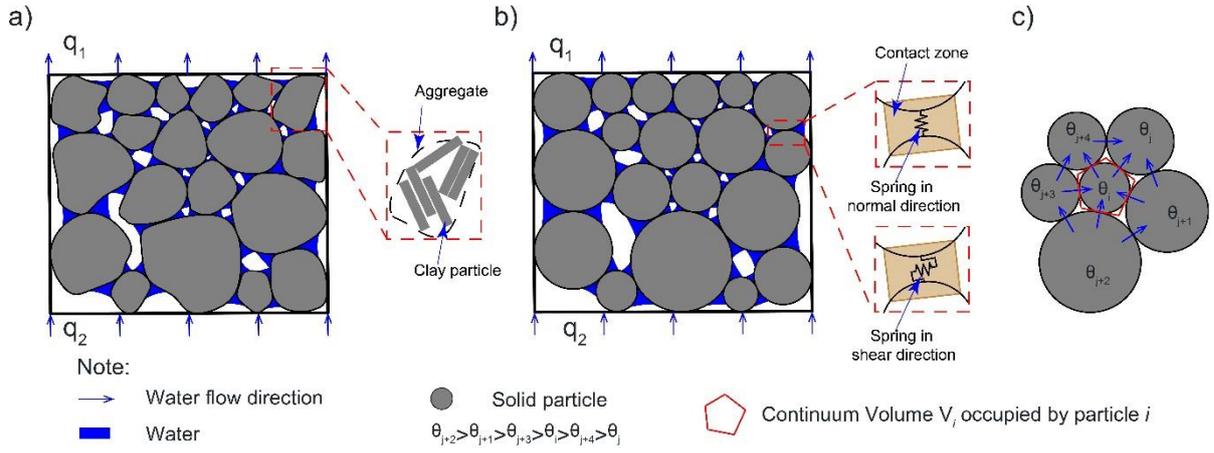

Fig. 1. Concepts of the hydro-mechanical DEM framework (a) clayey aggregates and boundary conditions; (b) idealised system in DEM model; (c) water flow directions between some particles.

## 2.2. Governing equations

### 2.2.1. Mass balance equations

Consider a representative particle that occupies an equivalent continuum volume $V$ as shown in Fig. 1(c), by applying the mass conservation law to this particle (or the equivalent continuum volume occupied by the particle), mass balance equations for solid and water phases can be written follows, respectively (Bui and Nguyen 2017):

$$\frac{d^s[\rho_s(1-n)]}{dt} = -\rho_s(1-n)\nabla \cdot \mathbf{V}_s \tag{1a}$$

$$\frac{d^w(\rho_w nS)}{dt} = -\rho_w nS(\nabla \cdot \mathbf{V}_w) \tag{1b}$$

where $\rho_s$ and $\rho_w$ are the intrinsic density of solid and water phases, respectively; $\mathbf{V}_s$ and $\mathbf{V}_w$ are the velocity of solid and water phases of the equivalent continuum volume, respectively; $n$ is the porosity; $S$ is the degree of saturation; and $\frac{d^s \pi}{dt}$ and $\frac{d^w \pi}{dt}$ denote the material derivative of field quantity $\pi$ on the solid and water phases, respectively.

By expanding Eq. (1a) and assuming that clay particles (not DEM particles) are incompressible, this equation can be simplified to:

$$\frac{1}{1-n}\frac{d^s n}{dt} = \nabla \cdot \mathbf{V}_s \tag{2}$$

By introducing the volumetric strain of the solid volume ($\varepsilon_{sk}$) into Eq. (2), the following equation can be obtained:

$$\frac{1}{1-n}\frac{d^s n}{dt} = \frac{d^s \varepsilon_{sk}}{dt} \tag{3}$$

Next, consider the material derivative of a field quantity $\pi$ on the solid and water phases, which can be written as follows:

$$\frac{d^s \pi}{dt} = \frac{\partial \pi}{\partial t} + \mathbf{V}_s \cdot \nabla \pi \quad \text{and} \quad \frac{d^w \pi}{dt} = \frac{\partial \pi}{\partial t} + \mathbf{V}_w \cdot \nabla \pi \tag{4}$$

These equations lead to the following relationship:

$$\frac{d^w \pi}{dt} = \frac{d^s \pi}{dt} + (\mathbf{V}_w - \mathbf{V}_s) \cdot \nabla \pi \tag{5}$$

Applying this relation to Eq. (1b) yields:

$$\frac{d^s (\rho_w n S)}{dt} + (\mathbf{V}_w - \mathbf{V}_s) \cdot \nabla (\rho_w n S) = -\rho_w n S (\nabla \cdot \mathbf{V}_w) \tag{6}$$

By rearranging this equation and introducing $\rho_w n S (\nabla \cdot \mathbf{V}_s) - \rho_w n S (\nabla \cdot \mathbf{V}_s)$ to the right side of this equation, Eq. (6) can be written in the following form:

$$\frac{d^s (\rho_w n S)}{dt} = -\nabla \cdot [\rho_w n S (\mathbf{V}_w - \mathbf{V}_s)] - \rho_w n S (\nabla \cdot \mathbf{V}_s) \tag{7}$$

The difference in velocities between the water and solid phases in Eq. (7) defines the water flux vector (**q**) through the deformed porous medium, which is often calculated by the following equation:

$$\mathbf{q} = n S (\mathbf{V}_w - \mathbf{V}_s) \tag{8}$$

Substituting Eq. (8) into Eq. (7) leads to the following equation:

$$\frac{d^s (\rho_w n S)}{dt} = -\nabla \cdot (\rho_w \mathbf{q}) - \rho_w n S (\nabla \cdot \mathbf{V}_s) \tag{9}$$

Expanding the left side of Eq. (9) and substituting Eq. (2) into Eq. (9), we obtain:

$$n S \frac{d^s \rho_w}{dt} + \rho_w S \frac{d^s n}{dt} + \rho_w n \frac{d^s S}{dt} = -\nabla \cdot (\rho_w \mathbf{q}) - \frac{\rho_w n S}{1 - n} \frac{d^s n}{dt} \tag{10}$$

By enforcing the incompressible of water to Eq. (10) and rearranging the equation, the following equation can be obtained:

$$\frac{S}{1 - n} \frac{d^s n}{dt} + n \frac{d^s S}{dt} = -\nabla \cdot \mathbf{q} \tag{11}$$

Substituting Eq. (3) into Eq. (11) we obtain:

$$S \frac{d^s \varepsilon_{sk}}{dt} + n \frac{d^s S}{dt} = -\nabla \cdot \mathbf{q} \tag{12}$$

In unsaturated soils, the capillary pressure ($p_c$) is a function of the degree of saturation (Bear and Cheng 2010), and this pressure can be calculated based on water pressure ($p_w$) and air pressure ($p_a$) using the following equation:

$$p_c = p_a - p_w \tag{13}$$

For unsaturated flow, air pressure is normally assumed to be zero (Bear and Cheng 2010), so the rate of change in the degree of saturation of a representative particle can be calculated as:

$$\frac{d^s S}{dt} = \frac{d^s S}{d^s p_c}\frac{d^s p_c}{dt} = \frac{d^s S}{d^s p_c}\left(-\frac{d^s p_w}{dt}\right) = \frac{d^s S}{d^s p_w}\left(\frac{d^s p_w}{dt}\right) = \frac{d^s S}{d^s h}\left(\frac{d^s h}{dt}\right) \tag{14}$$

where $h$ is the pressure head. Substituting Eq. (14) into Eq. (12), we obtain:

$$S\frac{d^s \varepsilon_{sk}}{dt} + n\frac{d^s S}{d^s h}\left(\frac{d^s h}{dt}\right) = -\nabla \cdot \mathbf{q} \tag{15}$$

The water flux vector can be also related to the water content from the following equations (Bear and Cheng 2010):

$$\mathbf{q} = -\mathbf{D}(\theta) \cdot \nabla \theta \tag{16}$$

where $\mathbf{D}(\theta)$ is the average diffusivity tensor between two particles or the macro-diffusivity tensor and $\theta$ is the volumetric water content.

Therefore, applying the chain rule to Eq. (15) to convert the derivative of water head to that of volumetric water content, the following mass balance equation considering the deformation of the solid skeleton can be obtained:

$$C_w\left(\frac{d^s \theta}{dt}\right) = -\left(\nabla \cdot \mathbf{q} + S\frac{d^s \varepsilon_{sk}}{dt}\right) \tag{17}$$

where $C_w = n\frac{d^s S}{d^s h}\frac{d^s h}{d^s \theta} = (1 - S\frac{d^s n}{d^s \theta})$.

Eq. (17) allows the volumetric water content of the representative particle to be calculated once the hydraulic constitutive relation (i.e. soil water retention curve and hydraulic conductivity curve) is known. This constitutive relation will be described later.

### 2.2.2. Momentum balance equations

Similar to the mass balance equations, the momentum balance equations can also be written for the solid and water phases. However, as the motions of the solid phase are described by the conventional DEM method, the momentum balance equations can be simply replaced by the translational and rotational motion equations. At the time step $t$, the motion of a representative particle (e.g. particle $i$) can be described as (Cundall and Strack 1979):

$$\frac{d^{s2}\mathbf{x}_i}{dt^2} = \frac{\mathbf{F}_i}{m_i} + \mathbf{g} \tag{18a}$$

$$\frac{d^s \boldsymbol{\omega}_i}{dt} = \frac{\mathbf{M}_i}{I_i} \tag{18b}$$

where $\mathbf{x}_i$ is the location of particle $i$; and $\boldsymbol{\omega}_i$ is angular velocity; $m_i$ is the total mass of particle $i$; $g$ is the acceleration of gravity; $I_i$ is the moment of inertia of particle $i$; $\mathbf{F}_i$ and $\mathbf{M}_i$ are the resultant force and moment vectors acting on particle $i$, respectively. The force and moment vectors can be calculated as follows:

$$\mathbf{F}_i = \sum_{c=1}^{N} F_{i,c}^n \mathbf{n} + \sum_{c=1}^{N} F_{i,c}^s \mathbf{s} + \mathbf{F}_i^{ext} + \mathbf{F}_i^{damp} \tag{19a}$$

$$\mathbf{M}_i = \sum_{c=1}^{N} \mathbf{F}_i \, \mathbf{l}_{i,c} + \mathbf{M}_i^{ext} \tag{19b}$$

where $\mathbf{F}_i^{ext}$ and $\mathbf{M}_i^{ext}$ are the external force and moment vectors, respectively, applying to the particle; $\mathbf{F}_i^{damp}$ is the damping force vector; $F_{i,c}^n$ and $F_{i,c}^s$ are the contact force of particle $i$ at contact $c$ in the normal ($\mathbf{n}$) and shear ($\mathbf{s}$) directions (Fig. 2). These forces are calculated using the following equations:

$$F_{i,c}^n = K_{c,n} u_{c,n}^e = k_{c,n} A_c u_{c,n}^e \tag{20a}$$

$$F_{i,c}^s = \sum K_{c,s} \Delta u_{c,s}^e = \sum k_{c,s} A_c \Delta u_{c,s}^e \tag{20b}$$

$$\mathbf{F}_i^{damp} = -\alpha |\mathbf{F}_i^*| sign(\mathbf{v}_i) \tag{20c}$$

where $u_{c,n}^e$ is the elastic displacement in the normal direction between two particles associated at contact $c$; $\Delta u_{c,s}^e$ is the elastic relative shear (tangential) displacement increment between two particles associated at contact $c$; $K_{c,n}$ and $K_{c,s}$ are the normal and shear contact stiffnesses, respectively; $k_{c,n}$ and $k_{c,s}$ are the normal and shear contact stiffnesses per initial unit area of the contact area at contact $c$, respectively; $\alpha$ is the damping coefficient; $\mathbf{F}_i^*$ is the resultant force vector of all contact forces, external force and gravity force acting on particle $i$; $\mathbf{v}_i$ is the translational or angular velocity of particle $i$; and $A_c$ is the contact area, which is equal to $\pi r^2$ for 3D and $2r$ for 2D simulation with $r$ being the minimum radius of two particles in contact.

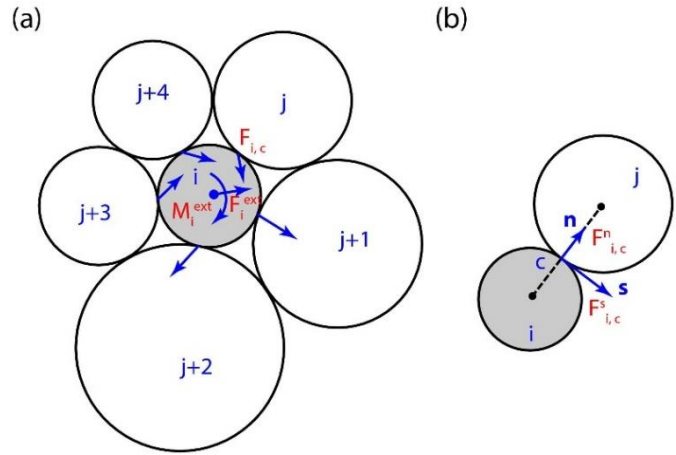

Fig. 2. (a) forces and moments acting on a particle; (b) forces at a contact.

As discussed above, the key assumption in this work is that the tensile and shear strengths of clay can be modelled by a set of cohesive-frictional springs, which can bear both tensile and shear loads. The tensile strength of the spring represents the hydraulic force causing by suction or molecular bonding force (i.e. Van der Waals forces) formed between two clay aggregates, while the shear strength of the spring represents the collective of frictional forces between two aggregates undergoing shearing. During the unsaturated stage of desiccation, the tensile and shear strengths of these springs could reduce to zero as the relative distance between two clay aggregates increases. To capture these essential mechanisms in the simplest way, the strength and stiffness of the springs can be modelled as a function of the volumetric water content ($\theta$), and the damage mechanics concept is employed to regulate the development of the tensile contact forces in DEM simulations during drying. A damage variable ($D_c$), whose magnitude can increase from 0 to 1, is introduced to the stiffnesses of normal and shear of the springs to account for their degradation as the relative distance between two clay aggregates increases:

$$k_{c,n} = \begin{cases} (1 - D_c)\, k_{c,n}^0(\theta) & \text{for tensile force} \\ k_{c,n}^0(\theta) & \text{for compression force} \end{cases} \tag{21a}$$

$$k_{c,s} = (1 - D_c) k_{c,s}^0(\theta) \tag{21b}$$

where the calculation of the damage parameter requires an appropriated constitutive model and the details of which will be presented in the next section; $k_{c,n}^0(\theta)$ and $k_{c,s}^0(\theta)$ are the initial spring stiffnesses of the contact $c$ and they are functions of the average water content $\theta$ between two clay aggregates; and $k_{c,n}$ and $k_{c,s}$ are the current spring stiffnesses corresponding the current state of the damage parameter $D_c$. These stiffnesses can be calculated from the effective modulus, $E^*(\theta)$ that is a function of water content and is further discussed in the subsequent section where the determination of model parameters is discussed, the normal-to-shear stiffness ratio ($k^*$) and the radii of particles $i$ and $j$ ($r_i$ and $r_j$) as follows (Itasca 2008):

$$k_{c,n}^0(\theta) = \frac{E^*(\theta)}{r_i + r_j} \tag{22a}$$

$$k_{c,s}^0(\theta) = k_{c,n}^0(\theta)/k^* \tag{22b}$$

Finally, to model the shrinkage of a clay aggregate due to water evaporation, the particle scaling concept (Guo et al. 2017) is adopted in this study. In particular, the radius of a clay aggregate $i$ at a given time step during the shrinkage process depends on the volumetric deformation of the equivalent continuum volume and is calculated as follows:

$$r_i^t = r_i^0 \sqrt[3]{(1 - \varepsilon_{sk,i}^t)} \tag{23}$$

where $r_i^0$ is the initial radius of particle $i$ ; $r_i^t$ is the radius of particle $i$ at time $t$; and $\varepsilon_{sk,i}^t$ is the volumetric strain of particle $i$ at time $t$, which is a function of water content and can be determined from the experiment. Further details on the calculation of volumetric shrinkage will be explained later.

### 2.3. Constitutive relations

#### *2.3.1. Hydraulic constitutive relation*

To complete the mass balance equations described in the previous section, the soil water retention curve (Fig. 3(a)) and hydraulic conductivity curve (Fig.3(b)) must be defined. These curves are used to link the volumetric water content (or the degree of saturation) and the hydraulic conductivity (or diffusivity) to the pressure head. In this work, the van Genuchten model (Van Genuchten 1980) is adopted to facilitate the simulation work:

$$S_e = \frac{S - S_r}{1 - S_r} = \frac{\theta - \theta_r}{\theta_s - \theta_r} = \frac{1}{[1 + (\alpha|h|)^n]^m} \tag{24a}$$

$$K(\theta) = K_s S_e^{\frac{1}{2}} \left[1 - \left(1 - S_e^{\frac{n}{n-1}}\right)^m\right]^2 \tag{24b}$$

where $S_e$ is the effective saturation; $S$ is the degree of saturation; $S_r$ is the residual degree of saturation; $\theta$ is the volumetric water content; $\theta_s$ is the saturated volumetric water content; $\theta_r$ is the residual water content; $h$ is the pressure head; $\alpha$, $n$ are curve fitting parameters; $m = 1 - \frac{1}{n}$; $K(\theta)$ is the hydraulic conductivity at the pressure head $h$ or the water content $\theta$, and $K_s$ is the saturated hydraulic conductivity.

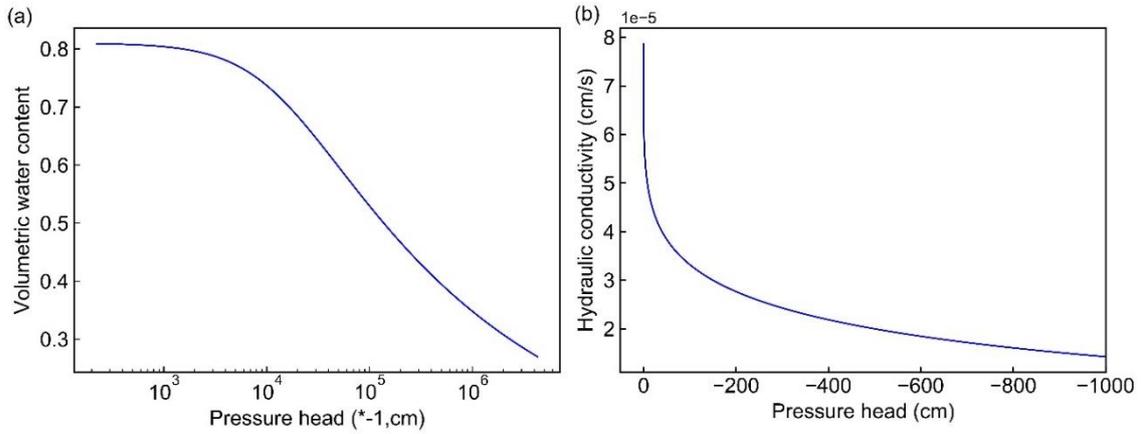

Fig. 3. Hydraulic constitutive relation: (a) soil water retention curve; (b) hydraulic conductivity

The diffusivity, $D(\theta)$, at a pressure head (or a water content) is then calculated as follows (Van Genuchten 1980):

$$D(\theta) = \frac{(1-m)K_s}{\alpha m(\theta_s - \theta_r)} S_e^{\frac{1}{2} - \frac{1}{m}} \left[\left(1 - S_e^{\frac{1}{m}}\right)^{-m} + \left(1 - S_e^{\frac{1}{m}}\right)^m - 2\right] \tag{25}$$

### 2.3.2. A cohesive-frictional contact model for the spring system

As discussed in the previous section, in this study, the clay is represented by an assembly of idealised spherical DEM particles of relatively large size (compared to clay particles), each of which represents a clay aggregate that is formed by several clay particles tied together due to inter-particle forces. Under this hypothesis, the interaction between two DEM particles is no longer the interaction between two individual clay particles, but the interaction between two clay aggregates subjected to all possible loading conditions, including compression, tension and shear. Dilatancy would occur between two DEM particles (or two clay aggregates) if they are subjected to compression shear load (similar to the direct shear box test, in which small particles rearrange and interact with each other). Accordingly, the conventional contact bond model (5 parameters) is no longer suitable, unless a sufficiently large number of particles is employed in DEM simulations to closely represent the micro-structure of the materials. Besides, previous studies have reported that clay soil experiences strength softening, plastic and ductile responses in three-point bending tests and direct tensile tests (Amarasiri et al. 2011, Nahlawi et al. 2004). This particular behaviour of clay cannot be properly captured using the conventional contact bond model (5 parameters), which is only suitable for highly brittle materials (i.e. perfectly brittle). Furthermore, during drying, cracks developed in clays are shown to undergo the combination of both shear and tensile failures (mixed-mode failure) (Wei et al. 2016). To properly describe these mechanisms, a cohesive-frictional cohesive framework (Nguyen et al. 2017a, Nguyen et al. 2017b) capable of regulating the mixed-mode failure is adopted in this work. The constitutive model starts with the decomposition of the relative displacement between two particles into elastic and plastic components:

$$\mathbf{u}_c = \mathbf{u}_c^e + \mathbf{u}_c^p \tag{26}$$

where $\mathbf{u}_c(u_{c,n}, u_{c,s})$ is the total relative displacement vector between two particles at a contact $c$; $\mathbf{u}^e(\delta u_{c,n}^e, \delta u_{c,s}^e)$ is the elastic vector component of the total relative displacement and $\mathbf{u}^p(u_{c,n}^p, u_{c,s}^p)$ is the plastic (or irreversible) vector component of the total relative displacement. These plastic relative displacements begin to develop after the bonding contact starts yielding and are the sum of the incremental plastic relative displacements in the normal and shear directions ($\delta u_{c,n}^p$ and $\delta u_{c,s}^p$), which can be calculated from a non-associated plastic potential function ($G$) by applying the flow rule of plasticity theory as follows:

$$\delta u_{c,n}^p = \delta \lambda \frac{G}{\sigma_{c,n}} \tag{27a}$$

$$\delta u_{c,s}^p = \delta \lambda \frac{G}{\sigma_{c,s}} \tag{27b}$$

where $\delta \lambda \geq 0$ is the plastic multiplier; $\sigma_{c,n}$ and $\sigma_{c,s}$ are the normal and shear stresses, respectively. The non-associated plastic potential function is:

$$G(\sigma_{c,n}, \sigma_{c,s}, D_c) \tag{28}$$
$$= \sigma_{c,s}^2 - \left[\sigma_{c,s}^0(\theta)\xi(D_c) - \sigma_{c,n}\tan\psi_c\right]^2$$
$$+ \left[\sigma_{c,s}^0(\theta)\xi(D_c) - \sigma_{c,t}^0(\theta)\xi(D_c)\tan\psi_c\right]^2 = 0$$

where $\xi(D_c) = 1 -$ with $D_c$ being the damage variable; $\sigma_{c,t}^0(\theta)$ and $\sigma_{c,s}^0(\theta)$ are the tensile strength and shear strength of the bonding contact, respectively, which are a function of the volumetric water content; and $\psi_c$ is the dilatancy angle of the contact. The damage variable is calculated from the plastic displacements in the normal and shear direction as follows:

$$D_c = 1 - e^{-\left(\frac{u_{c,n}^p}{u_{c,n}^0(\theta)} + \frac{u_{c,s}^p}{u_{c,s}^0(\theta)}\right)} \tag{29}$$

where $u_{c,n}^0(\theta)$ and $u_{c,s}^0(\theta)$ are softening parameters in the normal and shear directions, adopted to control the post-peak responses of the contact bond under the tensile and shear loading conditions.

More specifically, the smaller softening parameters, the greater reduction of the tensile and shear stresses after reaching their peak values.

The contact force at contact $c$ caused by these displacements is then calculated by using Equation (20), and thus the normal and shear stresses at this contact can be now obtained by using the following equations:

$$\sigma_{c,n} = \frac{F_c^n}{A_c} = \begin{cases} k_{c,n}^0(\theta)(1-D_c)(u_{c,n} - u_{c,n}^p) & \text{Tension} \\ k_{c,n}^0(\theta)(u_{c,n} - u_{c,n}^p) & \text{Compression} \end{cases} \quad (30a)$$

$$\sigma_{c,s} = \frac{F_s^n}{A_c} = k_{c,s}^0(\theta)(1-D_c)(u_{c,s} - u_{c,s}^p) \quad (30b)$$

Finally, the yield criterion regulating the yielding of the contact is defined as:

$$\begin{aligned} F(\sigma_{c,n}, \sigma_{c,s}, D_c) &= \sigma_{c,s}^2 - [\sigma_{c,s}^0(\theta)\xi(D_c) - \sigma_{c,n}\tan\varphi_c]^2 \\ &+ [\sigma_{c,s}^0(\theta)\xi(D_c) - \sigma_{c,t}^0(\theta)\xi(D_c)\tan\varphi_c]^2 = 0 \end{aligned} \quad (31)$$

where $\varphi_c$ is the friction angle of the contact.

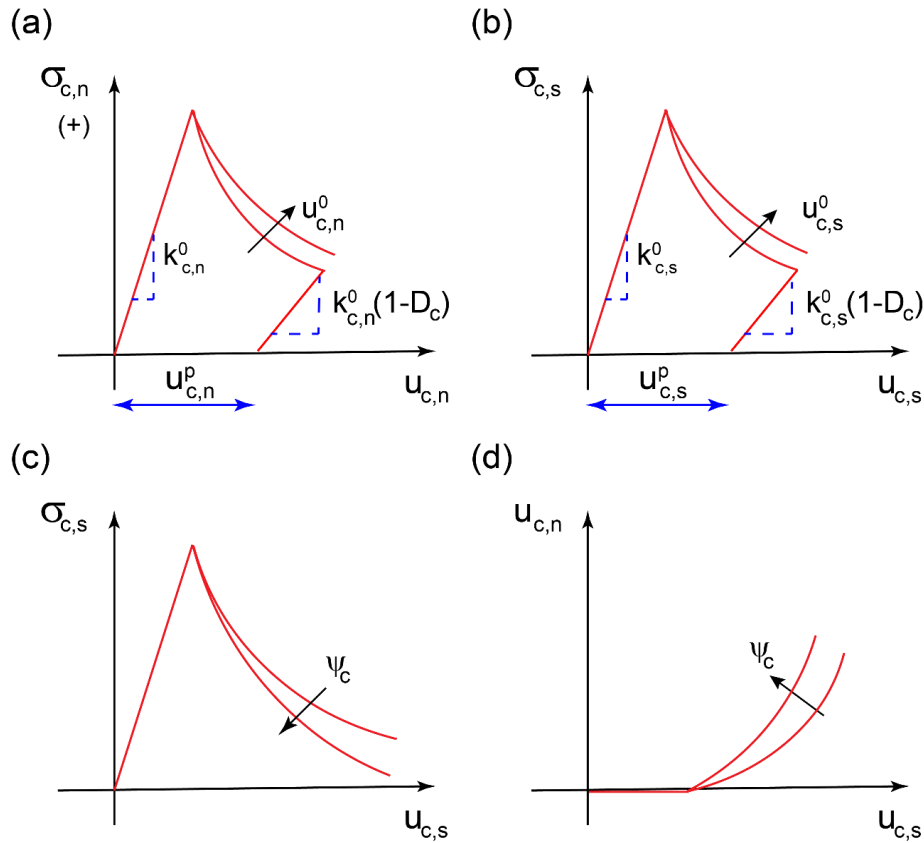

Fig. 4. Stress-displacement behaviour: (a) tensile stress; (b) shear stress; (c) shear stress changes with different dilatancy angles; (d) normal displacement changes with different dilatancy angles.

Figs. 4(a) and (b) shows the evolution of stresses at a contact under tensile and shear. As the normal and shear displacements increase, the tensile and shear stresses increase until reaching their peak values. With further loading, damage develops and the tensile and shear stresses gradually reduce in an exponential manner and reach zero value when $D_c = 1$. The effect of softening parameters on the responses of the tensile and shear stresses after experiencing peak values is also shown in these figures. After reaching these peak values, these stresses decrease more slowly as these softening parameters increase. Figs. 4(c) and (d) show the effect of dilatancy angle on the evolution of shear stress and normal displacement at a contact, respectively, when this contact is under constant compression stress and sheared at a constant speed. As the dilatancy angle increases, the shear stress drops faster after reaching its peak value, while the normal displacement increases faster.

The relationships between parameters governing the mechanical behaviour of the contact and the water content of the equivalent continuum volume will be presented in section of determination of model parameters.

### 2.4. Time integrations of the governing equations

In this section, the time integration of hydraulic and mechanical equations governed the hydro-mechanical response of clay undergoing desiccation cracking are presented.

#### 2.4.1. Time integration of the hydraulic equation

The hydraulic equation governs the unsaturated seepage flow from a particle $i$ to its neighbouring particles caused the difference in volumetric water content (or piezometric head) between two particles was established in section of mass balance equations and is rewritten as follows:

$$C_w \left( \frac{d^s \theta}{dt} \right) = - \left( \nabla \cdot \mathbf{q} + S \frac{d^s \varepsilon_{sk}}{dt} \right) \tag{32}$$

To derive the solution for this equation, consider a particle $i$ with inflows and outflows, as shown in Fig. 1(c). This particle has $N$ contacts with other particles, occupies an equivalent continuum volume $V_i$ and undergoes a volumetric deformation ($\varepsilon_{sk}$). The average water flux per unit volume occupied by particle $i$ can be presented as follows:

$$\langle \nabla \cdot \mathbf{q} \rangle_i = \frac{1}{V_i} \int \nabla \cdot \mathbf{q}_i \, dV \tag{33}$$

By applying the Gauss divergence theorem to the right-hand side of Eq. (33) to convert the volume integral to the surface integral and discretising this term, the following equation can be obtained:

$$\langle \nabla \cdot \mathbf{q} \rangle_i = \frac{1}{V_i} \sum_{j=1}^{N} \mathbf{q}_{ij} \cdot \mathbf{n}_{ij} \Delta S_{ij} = \frac{1}{V_i} \sum_{j=1}^{N} Q_{ij} \tag{34}$$

where $\Delta S_{ij}$ is the cross-section area of the flow between particle $i$ to particle $j$; $\mathbf{n}_{ij}$ is the outward unit normal vector of the surface $\Delta S_{ij}$; $\mathbf{q}_{ij}$ and $Q_{ij}$ are the water-flux vector and the discharge rate of water follow between particles $i$ and $j$, respectively, and can be calculated using Darcy's law as:

$$\mathbf{q}_{ij} = -d_{ij} \frac{\Delta \theta_{ij}}{L_{ij} \Delta S_{ij}} \mathbf{n}_{ij} \tag{35a}$$

$$Q_{ij} = \mathbf{q}_{ij} \cdot \mathbf{n}_{ij} \Delta S_{ij} = -d_{ij} \frac{\Delta \theta_{ij}}{L_{ij}} \tag{35b}$$

where $\Delta\theta_{ij}$ is the water content difference between particles $i$ and $j$; $L_{ij}$ is the distance between the centres of two particles and $d_{ij}$ is the inter-particle diffusivity calculated based on the micro-diffusivity of particles $i$ and $j$ ($d_i$ and $d_j$), which will be presented later.

Substituting Eqs. (34) and (35a) into Eq. (32) yields the following equation governing the water flow through particle $i$:

$$C_{w_i}\left(\frac{d^s\theta}{dt}\right)_i = \frac{1}{V_i}\sum_{j=1}^{N} d_{ij}\frac{\Delta\theta_{ij}}{L_{ij}} - S\frac{d^s\varepsilon_{sk}}{dt} \tag{36}$$

Using the forward-finite difference scheme, the volumetric water content $\theta_i^{t+\Delta t}$ of particle $i$ at a new time step can be estimated from the volumetric water content $\theta_i^t$ as follows:

$$\theta_i^{t+\Delta t} = \theta_i^t + \frac{\Delta t}{C_{w_i}}\left[\frac{1}{V_i}\sum_{j=1}^{N} d_{ij}\frac{\Delta\theta_{ij}}{L_{ij}} - S_i\frac{d^s\varepsilon_{sk}^i}{dt}\right] \tag{37}$$

To complete this equation, the inter-particle diffusivity between particles $i$ and $j$ is required and this can be calculated from the micro-diffusivity of particles $i$ and $j$ following the harmonic mean approach as suggested by Tran et al. (2019):

$$d_{ij} = \frac{2d_i d_j}{d_i + d_j} \tag{38}$$

where the micro-diffusivity of a particle $i$ can be obtained from its macro-diffusivity of soil $D(\theta_i)$ as follows:

$$d_i = \frac{3D(\theta_i)V_i}{\sum_{c=1}^{N} l_{ic}} \tag{39}$$

Finally, to ensure the stability as well as to guarantee the positive water content, the following criterion for the numerical time step is required (Tran et al. 2019):

$$\Delta t \leq \min\left[\frac{(\Delta z)^2 Fn}{D(\theta_i)}; \frac{\theta_i^t C_{w_i}}{\left[\frac{1}{V_i}\sum_{j=1}^{N} d_{ij}\frac{\theta_i^t}{L_{ij}} + S_i\frac{d^s\varepsilon_{sk}^i}{dt}\right]}\right] \tag{40}$$

where $\Delta z$ is the equivalent size of the continuum space occupied by particle $i$, which can be taken to be the diameter of particle $i$; and $Fn$ is the Peclet and Courant number, which is commonly selected as 0.5 (El-Kadi and Ling 1993, Haverkamp et al. 1977).

*2.4.2. Time integration of the motion equations*

To solve the momentum equations, the Leap-Frog time integration scheme is employed. First, the translational and rotational velocities of particle $i$ at time $(t + \Delta t/2)$ are calculated, then the location and rotational angle of this particle at time step $(t + \Delta t)$ are updated. More information can be found from literature (Cundall and Strack 1979, Itasca 2008). In addition, to ensure the numerical stability, the following numerical time step is required (Itasca 2008):

$$\Delta t \leq \min\left[\sqrt{m_i/k_i^{tran}}; \sqrt{I_i/k_i^{rot}}\right] \qquad (41)$$

where $k_i^{tran}$ and $k_i^{rot}$ are the translational and rotational stiffnesses calculated by considering the contribution of all particles contacting with particle $i$ (Itasca 2008).

Finally, the numerical time step required for modelling the hydro-mechanical behaviour of a soil medium consisting of many particles should be the smallest value obtained by Eqs. (40) and (41) computed for every particle.

## 3. Application of the proposed framework

In this section, the proposed hydro-mechanical framework is employed to predict the behaviour of clayey soil undergoing desiccation cracking observed in the laboratory tests. The experimental procedures to obtained desiccation cracking data and required soil parameters will be first presented and followed by the numerical simulations and analysis of desiccation cracking.

### 3.1. Outline of experiments and results

Shrinkage tests were conducted on slurry samples prepared by Werribee clay. This clay was taken from the Aquatic Centre site in Werribee, Victoria, Australia. This clay is highly reactive, with a dominance of Smectite. The main properties of this clay are presented in Table 1, while other properties can be found elsewhere (Shannon 2013, Tran et al. 2019). To prepare slurry samples for testing, Werribee clay was first dried, then crushed and sieved using a 0.425mm sieve. The clay passing through the sieve was then thoroughly mixed with distilled water at a water content of approximately 160%. This water content was higher than the liquid limit of the clay to ensure that the prepared mixture was easy to place in moulds. The resulting slurry was then poured into a mould and the mould with the mixture was vibrated for 5 minutes to eliminate trapped air. The sample was then covered and kept in a damp place for 24 hours to help moisture to uniformly distribute inside the sample. After this period, the sample was placed on a balance connected to a computer with a precision of 0.01 g, which allows measuring the water content at a precision of 0.01 %. A camera also connected to the computer was fixed above the sample to monitor the development of cracks on the surface. Photos and sample weights were recorded every 15 minutes. To measure the dying condition, a temperature and relative humidity sensor was placed close to the sample. Finally, the sample was dried in an air-conditioned room for several days until the weight of the sample reached a stable value. Fig. 5 shows the test set-up.

Table 1. Typical clay properties (after Shannon (2013)).

| Properties | Values |
|---|---|
| Liquid limit (%) | 127.0 |
| Plastic limit (%) | 26.0 |
| Plasticity index (%) | 101.0 |
| Linear shrinkage (%) | 22.0 |
| Shrinkage limit (%) | 13.0 |
| Soil classification | CH |
| Specific gravity | 2.66 |
| Hydraulic conductivity (m/s) | $2.2 \times 10^{-10}$ |

CH: Clay of high plasticity

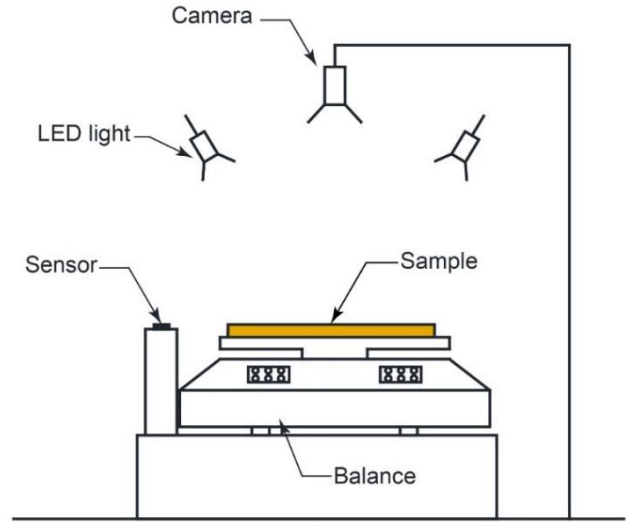

Fig. 5. Desiccation cracking test set-up.

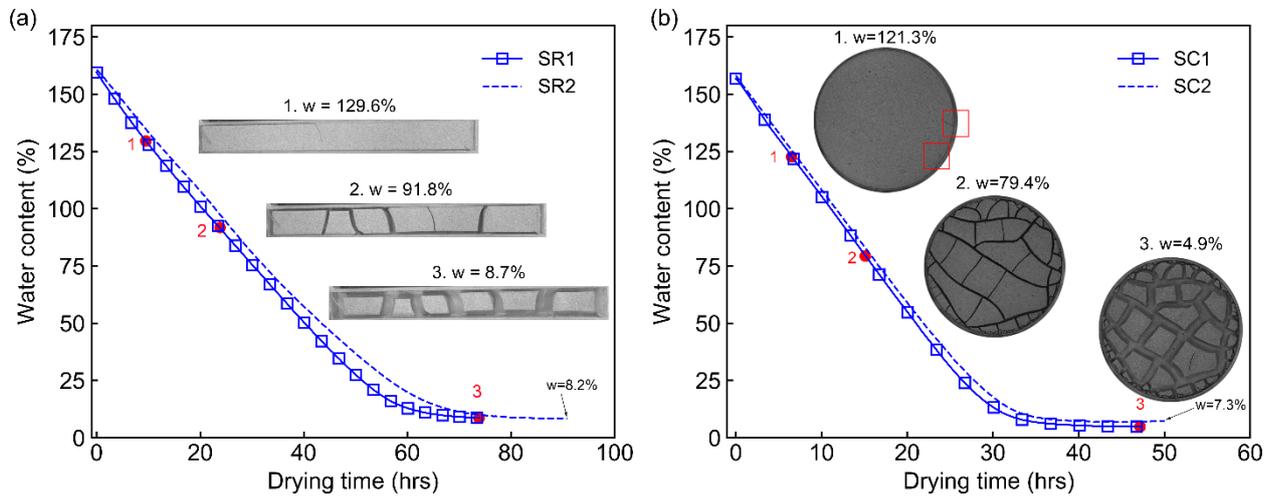

Fig. 6. Experimental results: (a) rectangular samples; (b) circular samples.

To validate the performance of the proposed numerical framework in predicting the development of complex cracking networks in clay soils, rectangular and circular samples were tested. The rectangular samples were 251 mm long, 25 mm wide and 12.4 mm deep, while the circular samples had a diameter of 140 mm and a thickness of 5 mm. For each shape, two identical samples were tested to assess the repeatability of the tests. In addition, a free shrinkage test was also conducted on a circular sample to determine numerical parameters for simulations. This sample was prepared using the same procedures employing to prepare other samples, except that grease was applied to all sides and the base of the mould to allow the sample to shrink freely. Table 2 shows the experimental program and testing conditions. All samples were tested in an air-conditioned room, in which only temperature was controlled. Since all samples were not tested at the same time, the range of relative humidity and temperature is different in these tests.

Table 2. Experimental program and testing conditions.

| Test ID | Mould shape | Initial water content (%) | Temperature range (°C) | Relative humidity range (%) | Grease applied to mould |
|---|---|---|---|---|---|
| SR1 | Rectangular | 159.6 | 21.9 – 24.4 | 29.3 – 42.5 | No |
| SR2 | Rectangular | 160.7 | 21.8 – 22.7 | 35.6 – 64.3 | No |
| SC1 | Circular | 156.7 | 20.9 – 23 | 27.5 – 36.6 | No |
| SC2 | Circular | 158.3 | 21.8 – 22.9 | 34.8 – 47.9 | No |
| Free_shrink | Circular | 155.3 | 22.4 – 24.6 | 39.4 – 46.2 | Yes |

Fig. 6 shows the typical experimental results of the shrinkage tests and its process can be briefly summarised as follows. As the desiccation process starts, the gravimetric water content of these samples decreases significantly. When the water content reaches around 133% and 122%, the first cracks start occurring in rectangular and circular samples, respectively. As the desiccation process continues, the gravimetric water content continues to decrease and more cracks appear in the samples. Once the water content of rectangular and circular samples reaches 92% and 80%, respectively, the number of cracks in the samples is almost stable. After this period, the water content continues to decrease and gradually approaches a stable value, while the existing cracks continue to widen. It is noticed that all cracks in the samples are formed before the gravimetric water content reaches its residual value, and the evolution of the gravimetric water content can be approximated as a linear relationship with time. Fig. 6 also shows that cracking network developed in the rectangular is quite simple with all cracks being relatively paralleled to the short edge of the sample, while the cracking pattern in the circular sample is very complex involving multiple cracks intersecting each other. These observations are consistent with previous studies on other clayey soils (Costa et al. 2013, Shin and Santamarina 2011, Tang et al. 2011).

### 3.2. Determination of model parameters

This section outlines the procedure to obtained model parameters and boundary conditions used in the simulations of soil desiccation cracking. These include volumetric shrinkage parameters, mechanical parameters, hydraulic parameters and hydraulic boundary conditions.

### 3.2.1. Volumetric shrinkage parameters

As discussed in the section where the momentum balance equations are presented, the volumetric shrinkage of clay samples due to water evaporation is modelled by reducing the radius of DEM particles (clay aggregates) following Eq. (23). The volumetric strain ($\epsilon_{sk}$) is proportional to the rate of water evaporation from the soil and thus can be formulated as a function of the gravimetric water content for the whole drying process (Kodikara and Choi 2006, Peron et al. 2009):

$$\epsilon_{sk} = \alpha \Delta w \tag{42}$$

where $\Delta w = w_0 - w_t$ is the change in gravimetric water content; $w_0$ is the initial gravimetric water content, $w_t$ is the gravimetric water content at time $t$; and $\alpha$ is the shrinkage coefficient, which can be obtained from the free shrinkage test.

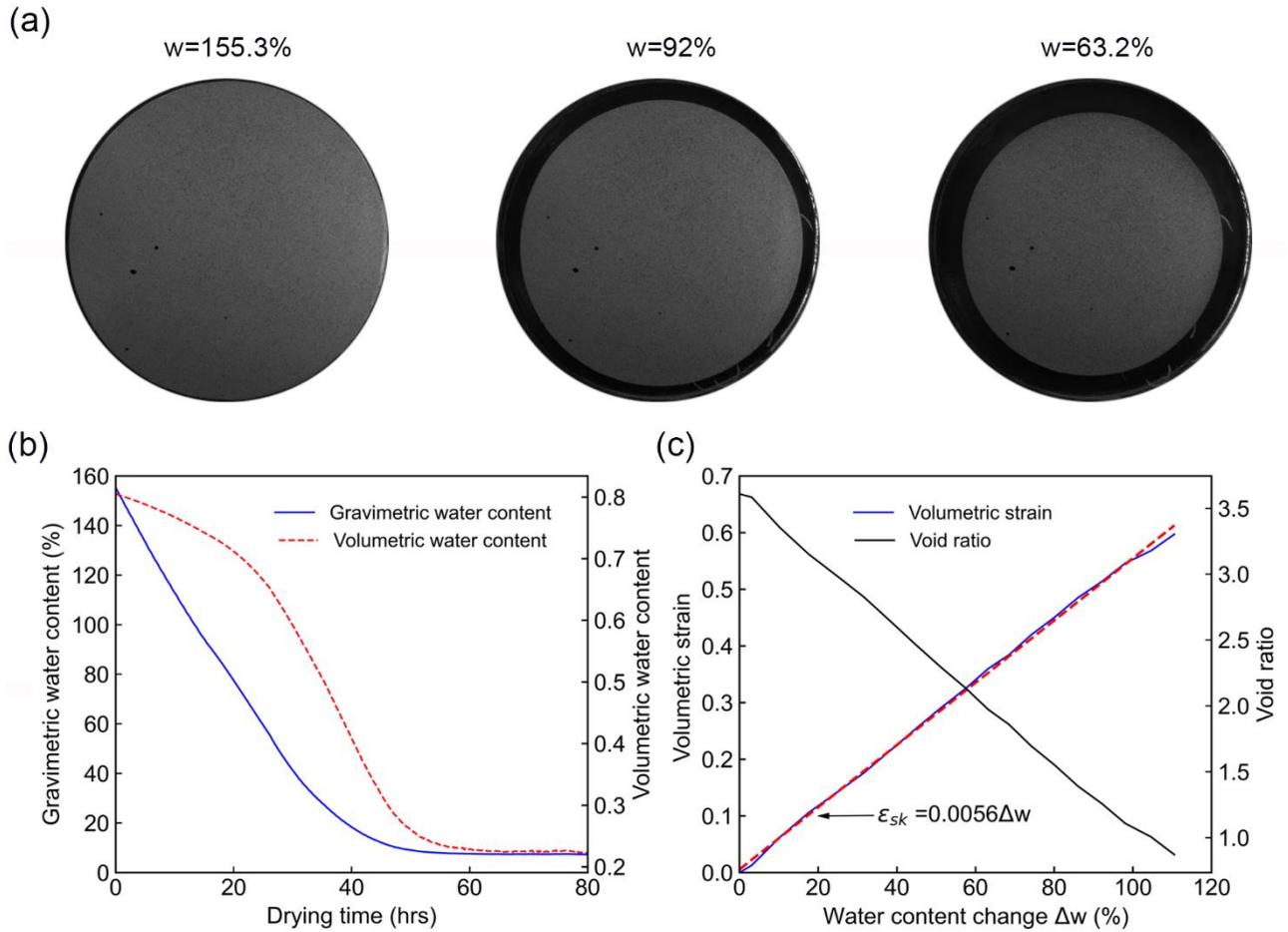

Fig. 7. Results of the free shrinkage test: (a) the shape of the sample; (b) changes in water content; (c) changes in volumetric strain and void ratio.

Fig. 7 shows the experimental results obtained from the free shrinkage test on the circular clay specimen. The sample undergoes uniform shrinkage in the radial direction, as demonstrated in Fig. 7(a). The gravimetric water content exhibits a linear reduction before reaching its residual value, while the volumetric water content follows the same trend at two noticeable differences in the evaporation rate before reaching the stable condition (Fig. 7(b)). The void ratio linearly decreases with an increase in the change in the gravimetric water content (Fig. 7(c)). Note that only a linear reduction part of the void ratio is plotted in this figure because the strain calculation of this sample is stopped due to soil curling occurs. The volumetric strain of the sample as a function of gravimetric water content was calculated prior to curling and cracking and is plotted in Fig. 7(c). It can be seen that Eq. (42) with a moisture shrinkage coefficient $\alpha = 0.0056$ fits very well to the experimental data, showing a linear relationship between the volumetric strain and the change of gravimetric water content. It is worth noting that the moisture shrinkage coefficient is dependent on the shape and the size of the sample. In this study, the moisture shrinkage coefficient was measured by conducting a free shrinkage test on the circular clay specimen, which has the same size as the desiccation cracking tests on circular samples, and this coefficient is used in the simulation of the circular sample. In addition, as the difference between the transverse and axial strains in rectangular samples is not significant (Peron et al. 2009), it is reasonable to use this coefficient in the simulation of the rectangular sample.

### *3.2.2. Parameters of the contact model*

The strength and other mechanical properties of clayey soil are functions of water content (Amarasiri et al. 2011, Mitchell and Soga 2005). Therefore, in this study, the micromechanical parameters required for the cohesive-frictional contact model between soil particles were obtained by simulating a series of three-point bending tests of samples prepared by Werribee clay at different water content reported by Amarasiri et al. (2011). Fig. 8(a) shows the numerical set-up, which is identical to that of the experiment reported by Amarasiri et al. (2011). The sample (140 mm long, 30 mm high, and 30 mm wide) with a notch of 10 mm in height at the middle is placed on two supporting plates located at a distance of 100 mm. This numerical sample was created by an assembly of 98943 DEM particles with radii varying uniformly from 0.45 mm to 0.7 mm. The sample was loaded by another plate located on the top surface at the centre of the sample. For each experimental specimen prepared at a given water content, a set of micromechanical parameters (i.e. tensile strength, effective modulus and softening parameters) was identified by fitting the numerical load-displacement curve to that of the experiment, while other properties (i.e. normal-to-shear stiffness ratio, friction coefficient, dilatancy angle and local damping) were chosen to be 2.0, 0.5, 0.1 and 0.7, respectively. Furthermore, the shear strength was assumed to be equal to the tensile strength, and the softening parameters were the same for both normal and shear directions (Nguyen et al. 2017).

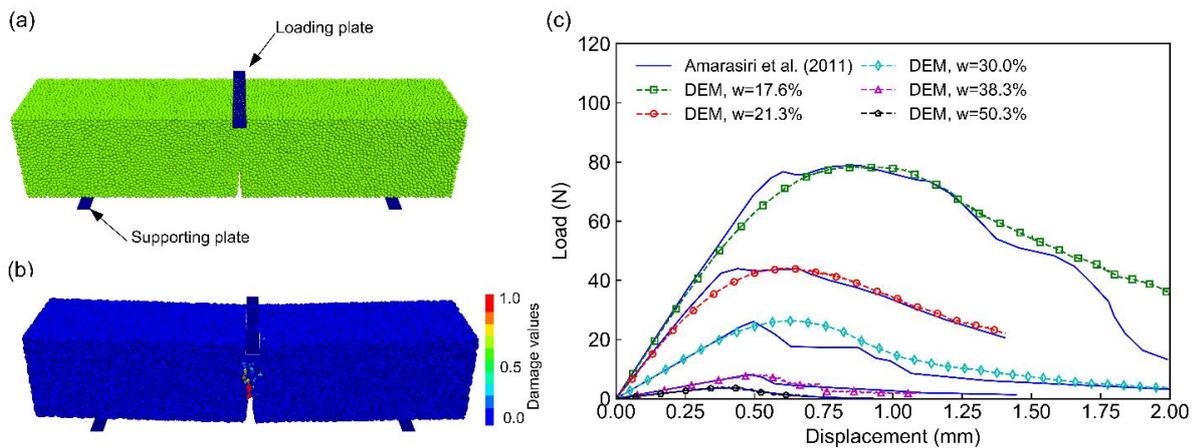

Fig. 8. Three-point bending test: (a) numerical set-up; (b) crack pattern at the end of the test (w=21.3%) ; (c) load-displacement curves.

Fig. 8(b) shows the crack pattern developed in the sample with a water content of 21.3% obtained from the numerical simulation. The crack begins from the notch tip and extends towards the top surface of the sample. A similar result was also reported in the experiment (Amarasiri et al. 2011), suggesting the proposed cohesive-frictional contact can capture the cracking development in clay soils. On the other hand, Fig. 8(c) shows the load-displacement curves of the three-point bending tests for samples with different gravimetric water content. It can be seen that, with an appropriate selection of the micro-constitutive parameters, the simulation can predict well the load-displacement curve obtained from the experiment. Furthermore, the proposed cohesive-frictional contact model can naturally capture the transition from ductile to brittle behaviour of clays as the water content changes, thanks to the coupled damage-plasticity of the model. From these simulations, the micromechanical constitutive parameters (as functions of the gravimetric water content) for the contact model between soil-soil particles can be identified and are plotted in Fig. 9. Finally, for contacts between soil-mould particles, the effective modulus, softening parameters, and strengths of soil–mould contacts were selected as 0.25, 0.5 and 0.5 of the values of soil–soil contacts, respectively (Guo et al. 2017, Sima et al. 2014).

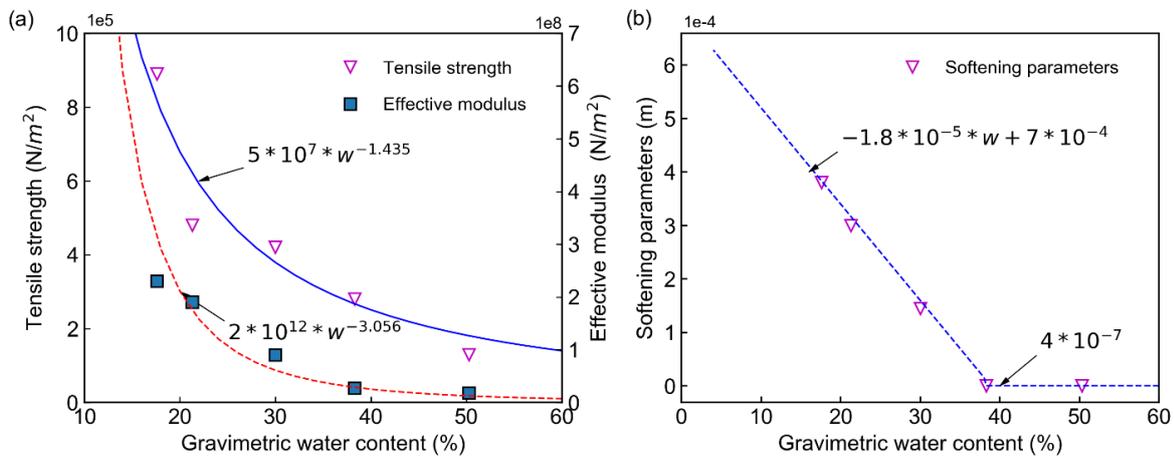

Fig. 9. Mechanical constitutive parameters of particle-to-particle contact.

### 3.2.3. Hydraulic parameters and hydraulic boundary conditions

The van Genuchten soil water retention model was adopted in this study to describe the transient behaviour of unsaturated seepage flow. This model requires four parameters, which can be obtained by fitting the van Genuchten model to the soil water retention curve obtained from the experiment. The soil water retention data reported by Nahlawi (2004) for Werribee clay (Fig. 10) were used to determine the van Genuchten model's parameters. The fitting parameters used for the van Genuchten model are $\theta_s = 0.81$, $\theta_r = 0.06$, $\alpha = 8.0772 \times 10^{-3}$ and $n = 1.2176$. In addition, the saturated hydraulic conductivity of 7.92×10$^{-7}$ m/hrs, which was previously reported in (Shannon 2013) for Werribee clay, is adopted in this study.

For hydraulic boundary conditions, water flux was imposed only on DEM particles located on the sample surface and crack surfaces. The flux was calculated based on the water evaporation rate obtained from the experiments. Fig. 11 shows the evolution of the water evaporation rate of the experimental samples. These data were calculated based on the initial area of the evaporative surface ($A_{ss0}$), which was 62.8 cm$^2$ and 154 cm$^2$ for the rectangular and circular samples, respectively. From this figure, the hydraulic boundary conditions for water evaporation rate were identified and applied to the sample surface as the water flux boundary condition for soil desiccation cracking simulations. On the other hand, the water flux boundary condition on cracking surfaces is expected to be much smaller than that of the sample surface and is taken to be 10% of the sample surface following the suggestion of Stirling et al. (2017). A more accurate measure of parameter requires sophisticated experiments to track the development of cracking depth and corresponding water flux, which are beyond the scope of this work. Accordingly, the water flux imposed on the sample surface ($E_{ss}^t$) and crack surfaces ($E_{cs}^t$) at a certain time $t$ during the simulation can be estimated from water evaporation rate ($E^t$) as follows:

$$E_{ss}^t = E^t A_{ss0}/(A_{ss}^t + 0.1 A_{cs}^t) \tag{43a}$$

$$E_{cs}^t = 0.1 E_{ss}^t \tag{43b}$$

where $A_{ss}^t$ is the sample surface area at time $t$, which can be the summation of the cross-section of particles on the sample surface and $A_{cs}^t$ is the total crack surface area at time $t$, which can be the summation of the cross-section of particles on the crack surfaces. It is noted that the detachment of the samples from the mould is also considered as the cracking surface.

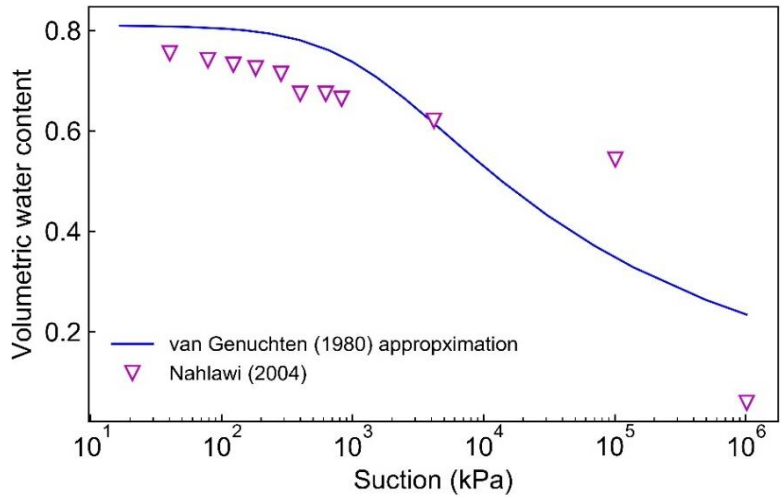

Fig. 10. Soil water retention curve of Werribee clay.

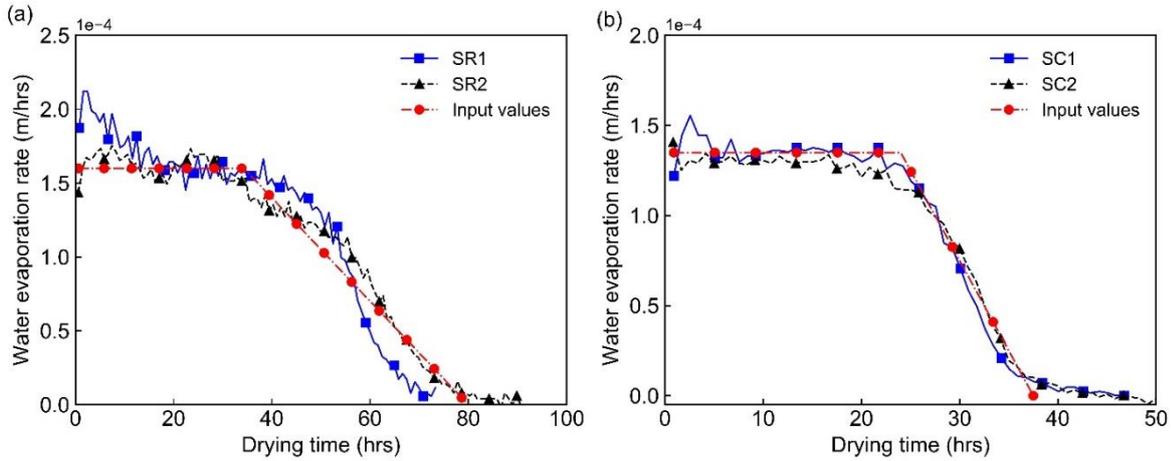

Fig. 11. Water evaporation rates of the experimental sample and numerical input values: (a) rectangular samples; (b) circular samples.

The water fluxes imposed on the sample and crack surfaces are then converted to the discharge rate for each DEM particle on theses surfaces as follows:

$$Q_i = qA_i \qquad (44)$$

where $Q_i$ is the discharge rate of particle $i$ on the boundary; $q$ is the water flux on the boundary ($q = E_{ss}^t$ for particles that are on the sample surfaces, while $q = E_{cs}^t$ for particles that are on the crack surfaces); $A_i$ is the area of the cross-section of particle $i$. This discharge rate $Q_i$ is used when the water content of particle $i$ is solved (using Eq. 35(b) and 37).

### 3.3. Numerical simulations of soil cracking experiments and insights

With all model parameters and boundary conditions identified in section of determination of model parameters, the proposed numerical framework can be now readily used to predict the desiccation cracking tests. Attention will be focused on assessing the predictive capability of the model in capturing the hydraulic (i.e. evolution of water content) and mechanical (i.e. evolution of cracking pattern) behaviour of the soil samples during desiccation.

#### *3.3.1. Numerical simulation of cracking of rectangular sample*

The numerical sample was created by randomly generating a predefined number of particles with their radii varying uniformly from 0.45 mm to 0.7 mm in a rectangular mould created by walls. This number of particles was calculated to create a numerical sample with the densest condition without overlapping to reduce the effect of stress-locking. All particles were then allowed to rearrange until reaching their equilibrium stage. During this process, the friction coefficient between all particles and walls was set as zero and all particles were not bonded together. By doing it, the numerical sample is homogenous and dense (Itasca 2008, Jiang et al. 2003). After that, all walls were deleted, and a total of 66,468 and 27,595 particles were assigned to be soil and boundary particles, respectively (Fig. 12). All particles were then bonded together and their initial properties including mechanical and hydraulic properties were assigned. Note that the same initial water content and hydraulic conductivity obtained from experiments were initially assigned to all soil particles. The drying process was then activated by initially imposing the flux boundary condition (Fig. 11) on the soil particles located on the sample surface and subsequently updating for soil particles on the crack surfaces when cracks develop in the numerical sample. The unsaturated flow information of each soil particle (e.g., water content, hydraulic conductivity) is automatically updated following the equations discussed in sections on time integrations and hydraulic constitutive relationship (e.g., using Eqs. 24, 25, 35(b) and 37). Other information including radii and locations of soil particles, and mechanical properties of contacts is then updated using information presented in Sections on momentum balance equations and determination of model parameters. During this process, the boundary particles were fixed in their position, while the soil particles can move. Fig. 13 shows a comparison of the changes in water content between the numerical and experimental samples during desiccation. The water content numerically predicted agrees well with the experimental counterpart, demonstrating that the proposed framework is able to capture unsaturated flow through deformed soil media.

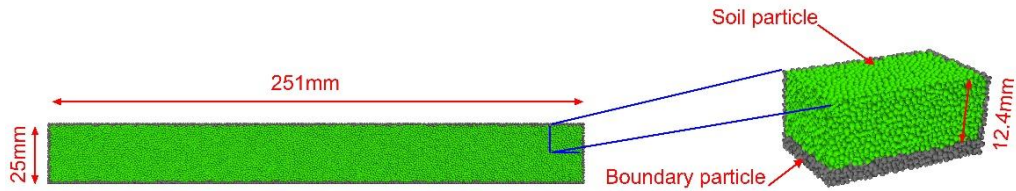

Fig. 12. Numerical rectangular sample.

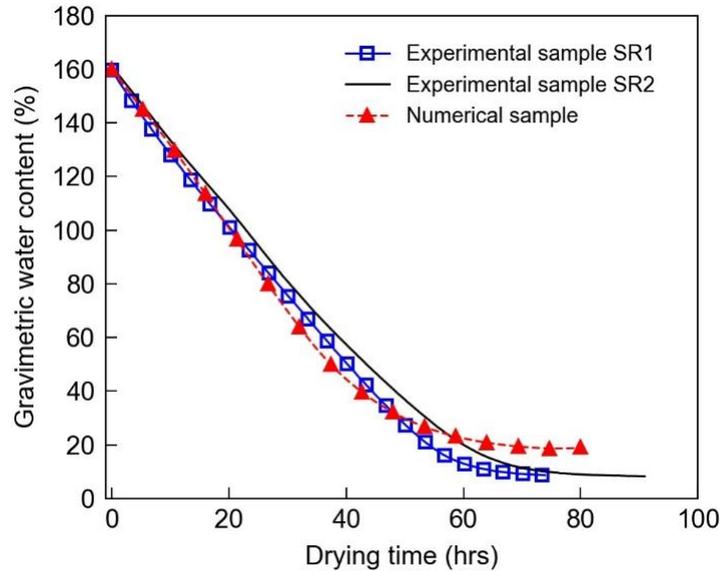

Fig. 13. Evolution of water content during simulation of the rectangular sample.

The development of cracking pattern in the numerical sample is shown in Fig. 14 and is compared to that observed in the experimental sample (SR1). To facilitate the visualisation of crack development in the numerical sample, soil particles on the sample surface are plotted in green, while the remaining soil particles are in red. It can be seen the initiation and development of cracks in the numerical sample reflect those observed in the experimental sample. At the beginning of the desiccation process, all samples start shrinking and detaching from the mould. The first crack then initiates from one long edge of the samples, propagates transversely and reaches the other long edge when the water content is approximately 127.1% and 124.2% for the experimental and numerical samples, respectively. As the desiccation process continues, the sample shrinks continuously, the first crack widens, while more cracks initiate and develop following the same behaviour as the first crack. When the water content is 117.6%, two more cracks appear in both samples. With further drying, no more cracks occur in the numerical sample and the width of the existing cracks continues to develop, whereas the widening process of existing cracks in the experimental sample progressed at a slower rate with two more cracks developed toward the end of the desiccation process. This could be attributed to the difference in the soil-mould interface between the experiment and simulation. The difference in the microstructure between the two samples could also contribute to this discrepancy. On the other hand, the prediction of water content at which the three first cracks occur in the numerical sample agrees well with the experiment counterpart. Although the number of cracks in the numerical sample is less than that in the experimental sample, the shape of the first three cracks (i.e. curved shape) is reasonably well captured by the simulation.

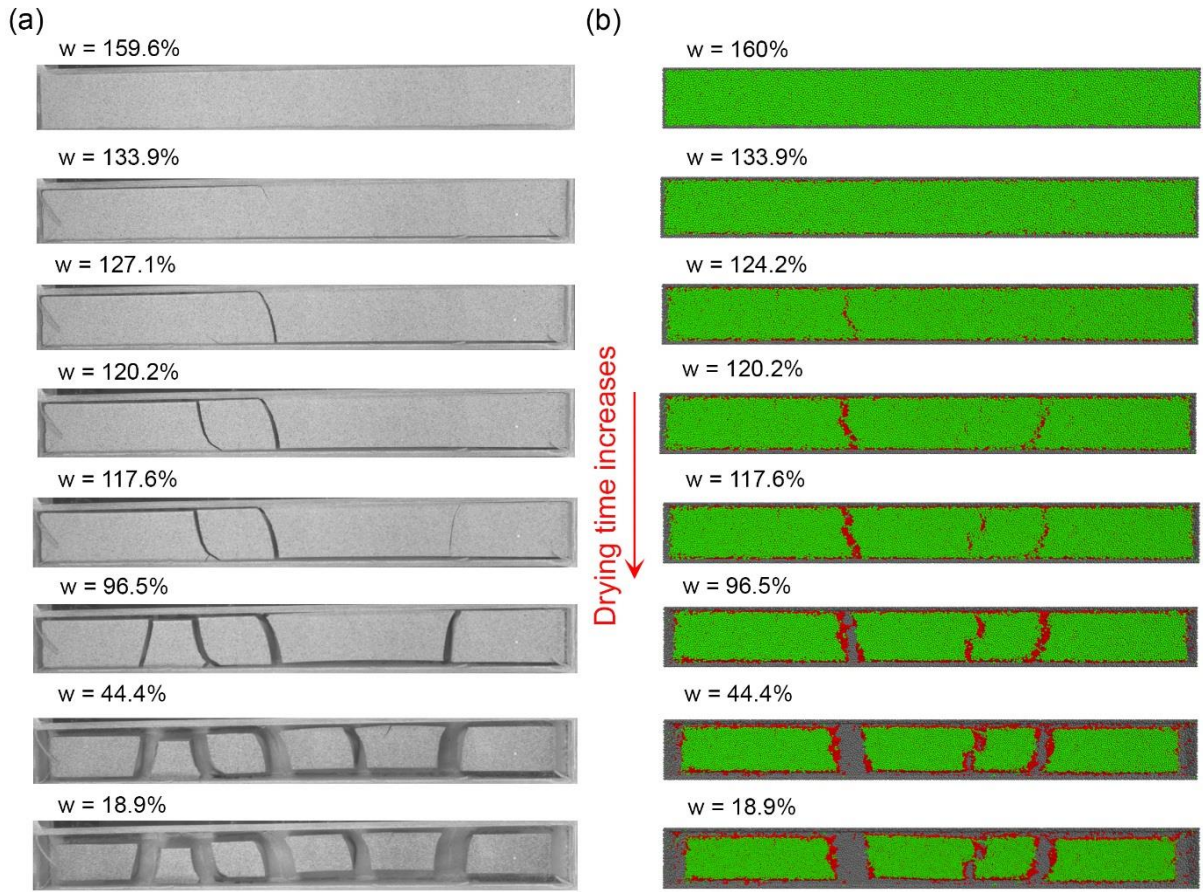

Fig. 14. Development of cracks in rectangular samples: (a) experimental sample SR1; (b) numerical sample.

In contrast to the results in Fig. 14, the comparison between the simulation against the experimental sample SR2 (Fig. 15) shows that the proposed numerical framework could predict fairly well the overall cracking pattern developed in the experiment. Nevertheless, the predicted water content at which cracks occurred appears to progress faster than that of the experimental counterpart (SR2). These differences are inevitable as they also occur in the experiment when two similar samples (SR1 and SR2) were dried under similar conditions (Figs. 14(a) and 15(a)). These may be attributed to the heterogeneous properties and the occurrence of flaws within samples. In this study, although the experimental samples were prepared from slurry clay with high water content (the initial gravimetric water content was 160%), these samples were not fully homogeneous, perhaps because air could be trapped inside or weak zones could be introduced during sample preparation. Besides, the heterogeneity of water content variation within the samples during the drying process of the sample causing different shrinkage strain within the sample may lead to the initiation of cracks. In particular, in these numerical samples, due to the difference in the microstructure, both strength and water content variation are not uniform within the samples. Besides, as it is impossible for the numerical samples to have the same microstructure with its experimental counterparts, numerical cracking processes can start at different moments with those in the experiments. Therefore, in the experimental samples in this study, cracks could occur earlier when the water content was high and more cracks might be developed. Note that the differences in water content when cracks appeared and in the number of cracks in the numerical and experimental samples have also been reported in previous studies when numerical results were compared to experimental results (Guo et al. 2017, Sima et al.

2014, Stirling et al. 2017, Vo et al. 2017). From the preceding discussion, it can be concluded that the proposed framework is capable of modelling soil desiccation cracking.

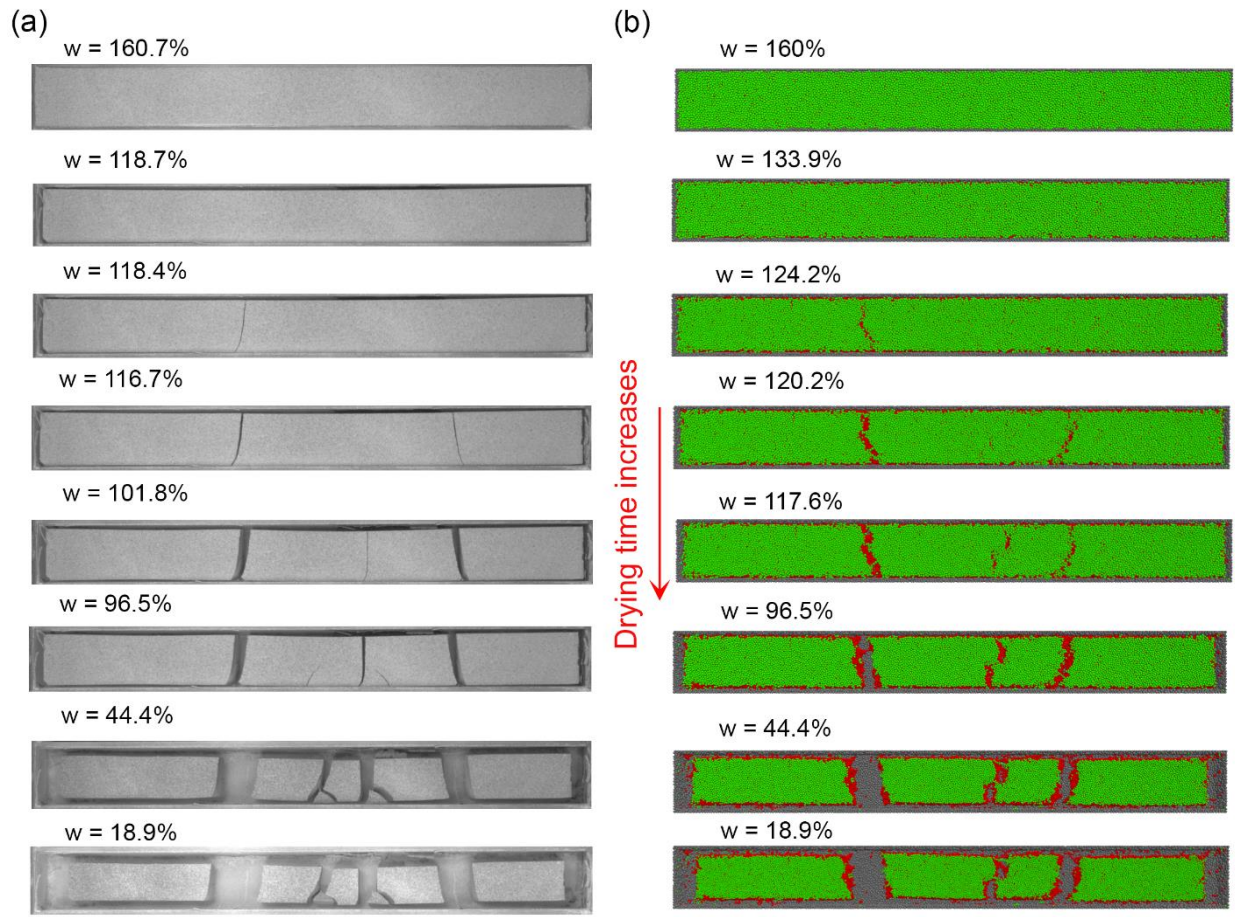

Fig. 15. Development of cracks in rectangular samples: (a) experimental sample SR2; (b) numerical sample.

As discussed for the methodology section, particles are bonded by a cohesive-frictional contact model representing the tensile strength clay due to suction and other inter-particle forces (e.g. van der Waal force) (Mitchell and Soga 2005). When the bonds are fully damaged ($D_c \approx 1$) and break, bond breakages, considered as micro-cracks, occur within the specimen. Macro-cracks will eventually form in the specimen from connected micro-cracks and progress with the evaporation of moisture. Fig. 16 shows the development and distribution of micro-cracks in the numerical sample. In this figure, small circles represent contact bonds that have been broken (i.e. a micro-cracks) in the specimen. To facilitate discussion, these micro-cracks are also classified into soil–soil micro-cracks and soil–boundary micro-cracks, which represent breakage of soil–soil and soil–boundary contact bonds, respectively. It can be seen that, as water evaporates, the sample begins to shrink and micro-cracks develop. At the water content of 133.9%, micro-cracks develop around the sides of the sample, mostly consisting of soil–boundary micro-cracks, indicating the detachment of the sample from the sides of the mould. As water content decreases, micro-cracks continue to develop. When the water content is 124.2%, soil–boundary micro-cracks appear at the base and around the two ends of the sample, while soil–soil micro-cracks appear within the sample, concentrated along a line indicating the formation of the first macro-crack. After the formation of the first macro-crack, more soil–boundary micro-cracks appear at the base and around the macro cracks, while more soil–soil micro-cracks are concentrated at some locations within the sample. The concentration of soil–boundary micro-cracks at the base around the macro cracks indicates the horizontal shrinkage of soil particle along the sample length and the strain concentration at the crack lips. When water content is 117.6%, soil–soil micro-cracks are concentrated along three lines forming three macro-cracks. After this period, soil–soil micro-cracks appear randomly inside the sample, while soil–boundary micro-cracks concentrate at the base, indicating that the shear strength resistance at the base from this stage onward mainly comes from the sliding friction of the material. It is worth noting that all existing works on DEM modelling of desiccation cracking in clayey soils could not simulate the moisture evaporation occurring at the particle scale. As a result, they could not properly capture the above localised failure mechanisms occurring during desiccation cracking in clay soils (Guo et al. 2017, Sima et al. 2014).

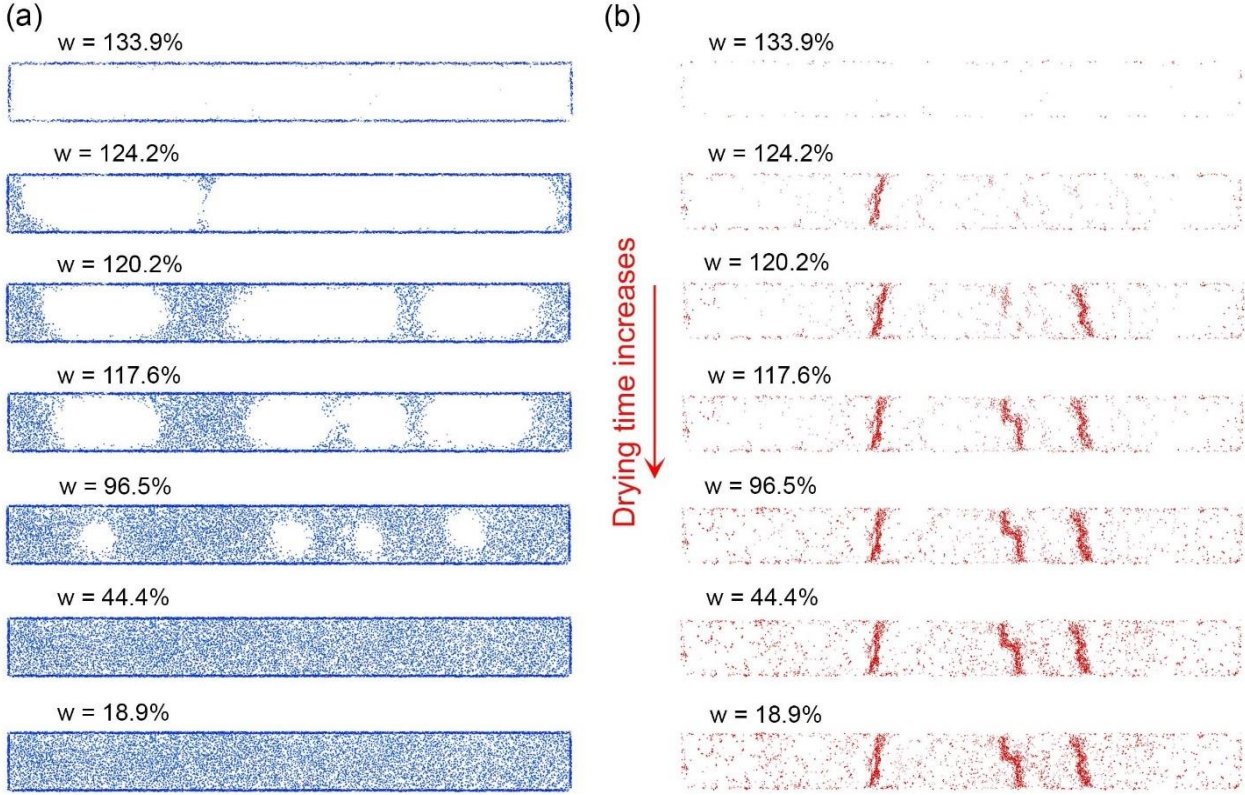

Fig. 16. Development of micro-cracks in the rectangular sample: (a) soil-boundary micro-cracks; (b) soil–soil micro-cracks.

The quantitative measure of the development of micro-cracks in the numerical sample during desiccation is plotted in Fig. 17. Soil–boundary micro-cracks begin to occur after 3.7 hours of drying (i.e. 149.4% water content), whereas soil–soil micro-cracks begin to appear after 5.8 hours of drying (i.e. 143.6% water content). The number of micro-cracks then increases significantly and gradually approaches stable values. After 46.1 hours of drying (i.e. 41.5% water content), the number of new micro-cracks of both types is negligible. During drying, the number of soil–boundary micro-cracks is much greater than that of soil–soil micro-cracks. For example, after 20 hours of drying (i.e. 100.7% water content), there are 11883 soil–boundary micro-cracks comparing to 2896 soil–soil micro-cracks.

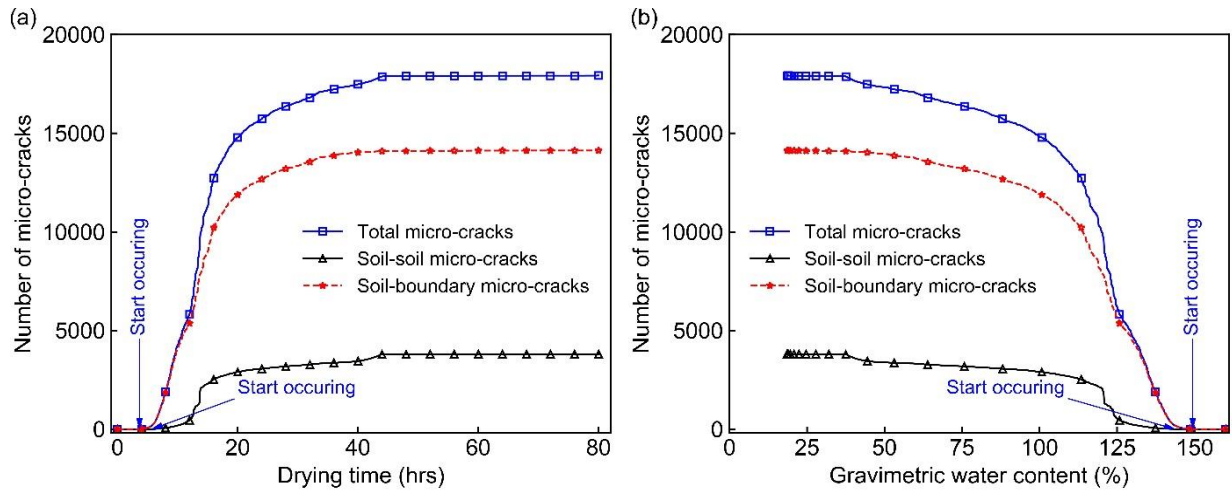

Fig. 17. Evolution in the number of micro-cracks during simulation of the rectangular sample: (a) the number of micro-cracks against drying time; (b) the number of micro-cracks against gravimetric water content.

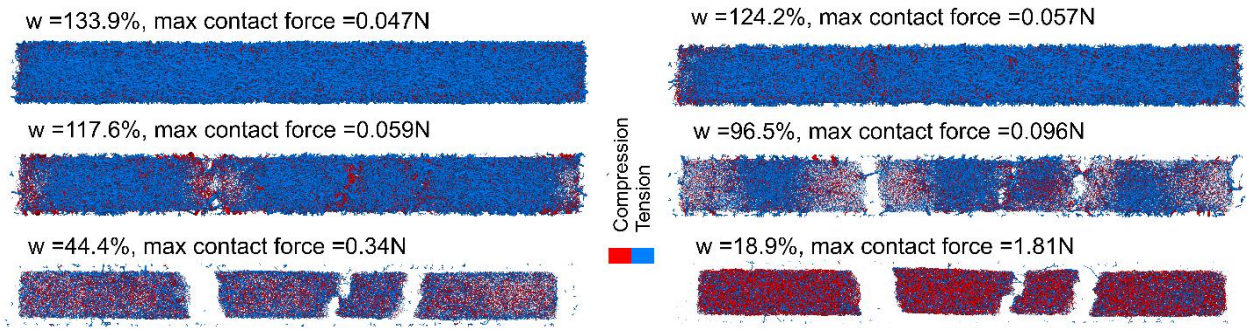

Fig. 18. Changes in the contact force network in the numerical rectangular sample during drying.

Fig. 18 shows the evolutions of the contact force network within the numerical sample during desiccation. At the beginning of the drying process when the water content is very high (e.g. 133.9%), tensile force dominates the compressive force and distributes relatively uniform across the sample. The preferential direction of the tensile force is the long edge of the sample. With further drying, micro-cracks develop, leading to the occurrence of macro-cracks that subdivide the sample into several sub-soil blocks. After this stage, the tensile force tends to concentrate in the middle of these sub-soil blocks without changing its preferential direction, while the compressive force develops mainly at crack lips. In addition, the compressive force gradually dominates the tensile force. Towards the end of the drying process when the water content of the sample is very low (e.g. 18.9%), the compressive force dominates the entire sample. The above evolution of the contact force network observed in the numerical sample provides further understanding of why cracks occur closely at the middle of the sample or of sub-soil blocks and their direction is somehow perpendicular to the long edge of the sample. This evolution also explains why the compressive strain mainly occurs in crack lips and tensile strain occurs in the middle of samples or soil blocks, which have been reported in recent experimental studies using the digital image correlation technique (TRAN 2019, Wang et al. 2018).

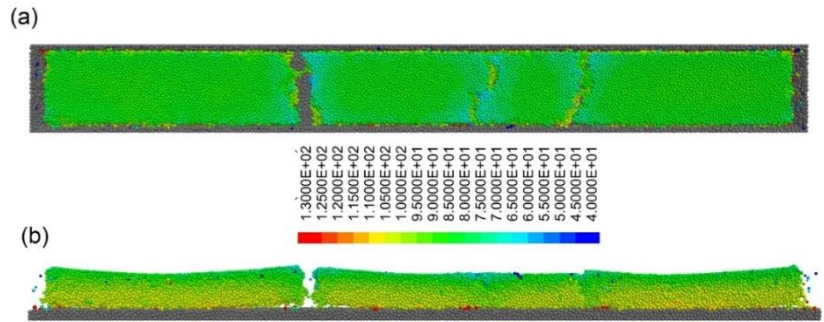

Fig. 19. Distribution of gravimetric water content (%) in the numerical sample with a rectangular shape at the average water content of 96.5%: (a) top view; (b) side view.

The distribution of gravimetric water content in the numerical sample at the average water content of 96.5% is plotted in Fig. 19. It can be observed that the water content distributes relatively uniform across the sample surface, except at/on crack lips where the moisture evaporation occurs and thus the water content is smaller. On the other hand, the distribution of the water content in the middle section of the sub-soil blocks increased toward the depth and this simulated result is consistent with experiments reported in the literature. For example, Nahlawi and Kodikara (2006) conducted desiccation cracking tests of slurry clayey soils and measured the water content of the top half and bottom half of samples. They reported that the top half always had less water content than the bottom half in testing, which can be also visualised from Fig. 19. This agreement again confirms that the proposed framework can capture the hydraulic behaviour of samples during desiccation. Furthermore, concave-up curling occurs in the simulation (Fig. 19), indicating that the increase in the number of soil–boundary micro-cracks after the appearance of all macro-cracks could also be attributed to the occurrence of concave-up curling.

*3.3.2. Numerical simulation of cracking of circular sample*

In this second test, the proposed hydro-mechanical DEM framework was used to predict cracking development in circular clay samples. Different from the rectangular samples where cracks are more and less developed and propagated parallel to the short edge of the samples, replicating mode-I fracture mechanism, the circular samples facilitate the development of mixed-mode fracture and allowing more complex fracture pattern to be developed. The numerical sample was created following the same procedure with particle radii used to prepare the rectangular sample presented in the section of numerical simulation of rectangular sample. It requires 64751 soil particles and 38298 boundary particles to generate a circular sample with an internal diameter of 140 mm and a thickness of 5 mm, respectively. Subsequently, the same initial water content and hydraulic conductivity obtained from experiments were initially assigned to every DEM particle. The hydraulic boundary condition shown in Fig. 11(b) was then applied to the circular sample to simulate the drying process, and the same set of soil parameters presented in the previous section were adopted in the simulation.

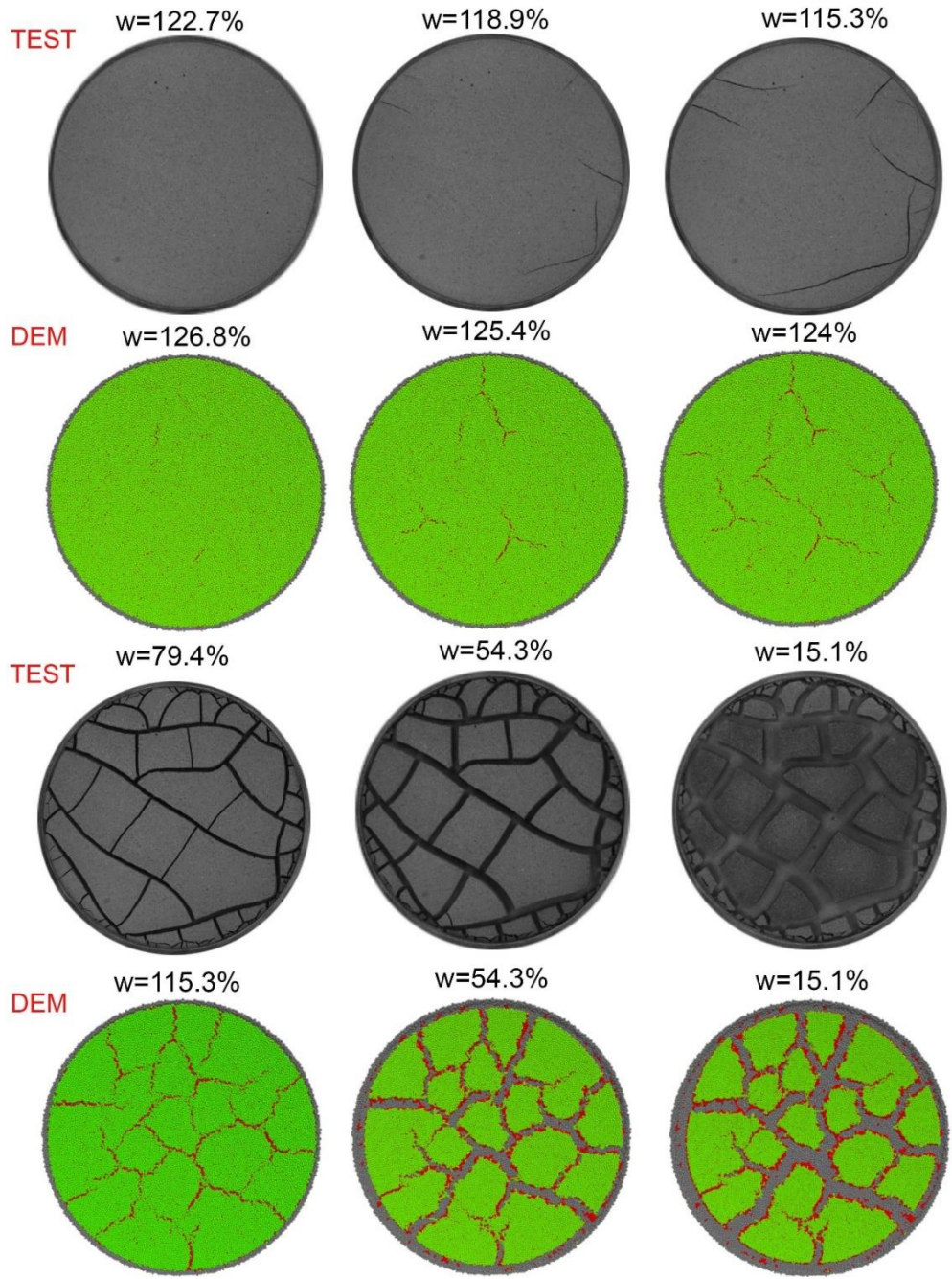

Fig. 20. Development of cracks in numerical sample and experimental sample (SC1).

Fig. 20 shows a comparison between the simulation and experiment (SC1) for the development of cracking network in the soil specimens. It can be observed that both experimental and numerical samples show similar crack development process including (1) cracks initiate and propagate radially or circumferentially towards the centre or boundary of the samples to subdivide the samples into smaller soil block; (2) more cracks begin at existing cracks and propagate until reaching other cracks while existing cracks become wider; (3) no more crack exists, while existing cracks continue to open. In addition, the crack pattern occurring in the experimental and numerical samples is similar. This pattern consists of many curved lines intersecting with each other at angles varying from more than $60^0$ to around $120^0$. However, cracks in the numerical sample occur and develop faster than cracks in the experimental sample. For example, the water content when first cracks occur is 126.8% and 122.7% for the numerical and experimental samples, respectively. This difference is inevitable because the distribution of heterogeneous properties and flaws within the experimental sample cannot be exactly modelled in the numerical sample. Besides, there is a difference in the locations of first cracks in the experimental and numerical samples. In the experimental sample, first cracks initiate and propagate from the boundary of the sample, while in the numerical sample, first cracks begin and develop somehow in the centre of the sample. This could be attributed to the difference in the soil-mould interface between the experimental and numerical samples. In the experimental sample, air could be possibly trapped at the interface between the soil sample and the mould during the sample preparation process, leading to the occurrence of flaws at the interface boundaries. Therefore, in the experiment, first cracks could start at these flaws and propagate. On the other hand, in the numerical sample first cracks initiate at locations with high stress concentration and these locations appear to be at the centre of the sample as shown in Fig. 23. This information was also reported in litterature (Costa et al. 2013, Peron et al. 2013). In fact, this difference is very common and has been reported in literature (Guo et al. 2017, Sima et al. 2014). Further investigations are required to improve the proposed approach so it could more quantitatively capture the behaviours of clay specimens observed in the experiment.

The development and distribution of micro-cracks in the numerical sample with circular shape is plotted in Fig. 21. To increase the visualisation, the size of soil-soil micro-cracks when water content is 128.2% and 126.8% is plotted bigger than that for other cases of water content. Before the occurrence of first macro-cracks, when the water content is very high (i.e. 128.2%), soil–boundary micro-cracks occur around the side of the sample, indicating the trend of the detachment of the sample from the mould, while soil–soil micro-cracking occurs randomly inside the sample, indicating the trend of the formation of macro-cracks. This second feature could occur in the circular sample as forces at every contact inside the sample could develop equally in both horizontal and lateral directions, and thus bonds at contacts could break due to mixed-mode failure. With further drying, soil–boundary micro-cracks continue to appear at the sample side, while soil–soil micro-cracks concentrate inside the sample, forming first macro-cracks. After the formation of first macro-cracks (e.g. w=124%), soil–boundary micro-cracks start to occur at the bottom surface of the sample, indicating the shrinkage deformation of the sample and the concentration of shrinkage strain at crack lips, while soil–soil micro-cracks continue to concentrate at specific locations to form macro-cracks. After water content reaches 115.3%, more soil–boundary micro-cracks continue to appear at the bottom surface of the sample, while soil–soil micro-cracks appear randomly within the sample and thus no more macro-crack forms. This process continues until the end of the test.

To have a quantitative view on the occurrence of micro-cracks during drying, changes in the number of micro-cracks during the test are plotted in Fig. 22. Similar to that of the rectangular sample, soil–boundary and soil–soil micro-cracks start occurring very early, just after approximately 3.4 hours drying, corresponding to a water content of approximately 135%. After this period, more micro-cracks (soil–boundary and soil–soil micro-cracks) continue to occur until the sample is dried approximately 20 hours. After this period, the water content of the sample is very high (e.g. 47%) and water is still evaporating, but only a few micro-cracks appear. In addition, during drying, the number of soil–boundary micro-cracks is much greater than that of soil–soil micro-cracks except the early stage of the drying process (up to 6.3 hours drying). At this stage, the number of soil–soil micro-cracks is slightly higher than that of soil–boundary micro-cracks, indicating that the sample detachment from the mould and the occurrence of macro-cracks may occur at the similar time to gradually form a complex crack pattern.

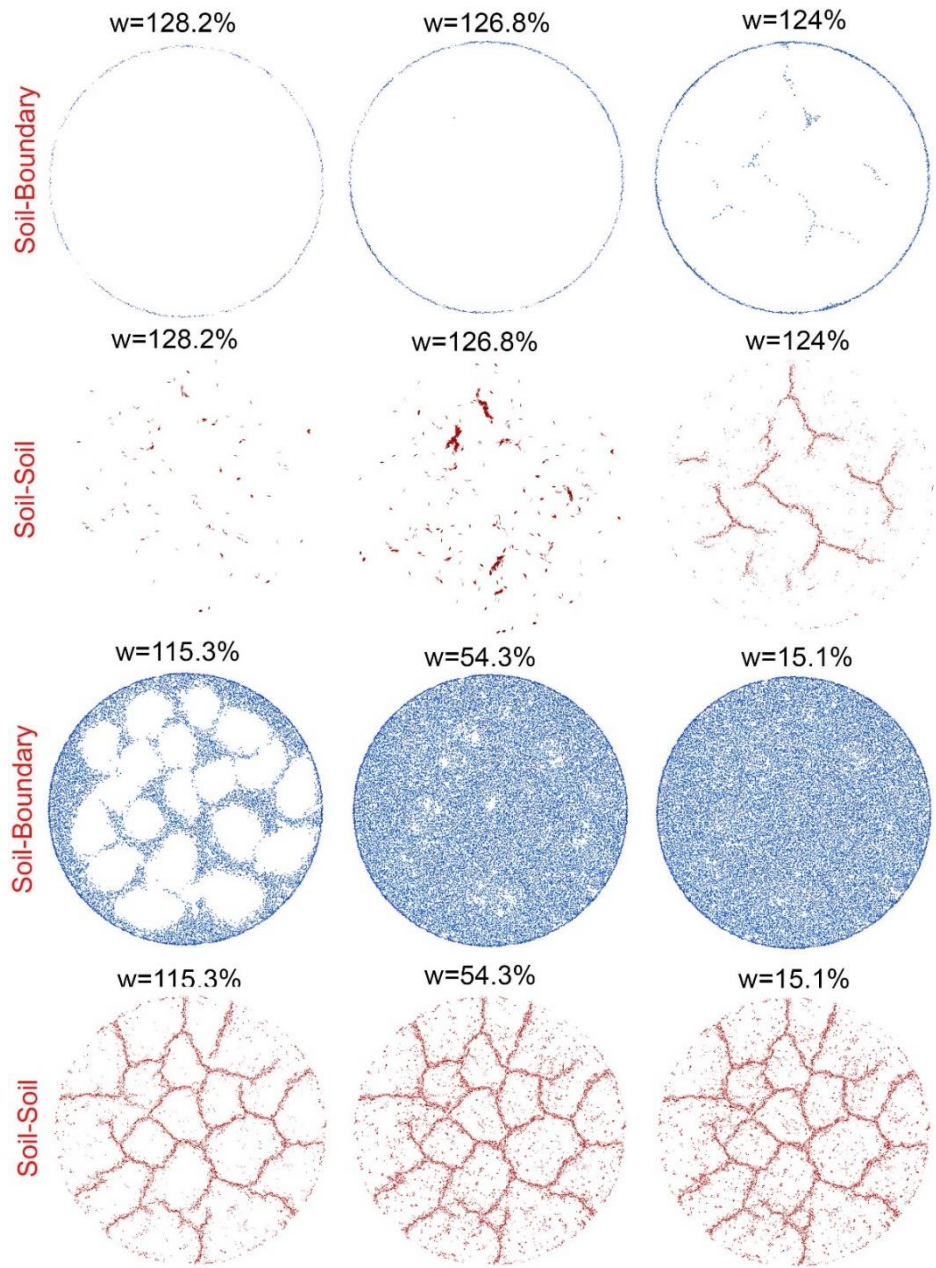

Fig. 21. Development of micro-cracks in the circular sample: (a) soil-boundary micro-cracks; (b) soil–soil micro-cracks.

Fig. 23 shows the evolution of the contact force network in the circular sample during drying. When the water content is high (i.e. 128.2%) and no macro-crack occurs, contact forces distribute uniformly in the sample, and most forces are tensile force, but the preferential direction of tensile force is not clear. With the occurrence of some macro-cracks, contact forces tend to redistribute. However, tensile force still dominates and tends to concentrate at crack tips or in the middle of the soil blocks if they form, while compressive force gradually exists more, especially at crack lips. The preferential direction of compressive force is somehow perpendicular to the cracks, indicating the development of compressive strain at these areas. In addition, around crack lips, tensile force tends to be parallel to the cracks, while around crack tips, this force is not perpendicular to the cracks. These features indicate that the cracks may continue to propagate due to mixed-mode failure and occur at the middle of the soil blocks. At the end of the drying process, the water content is lower (w=15.1%) the crack network is stable. The network becomes fairly isotropic and homogeneous. No clear pattern of the distribution of tensile and compressive forces is observed.

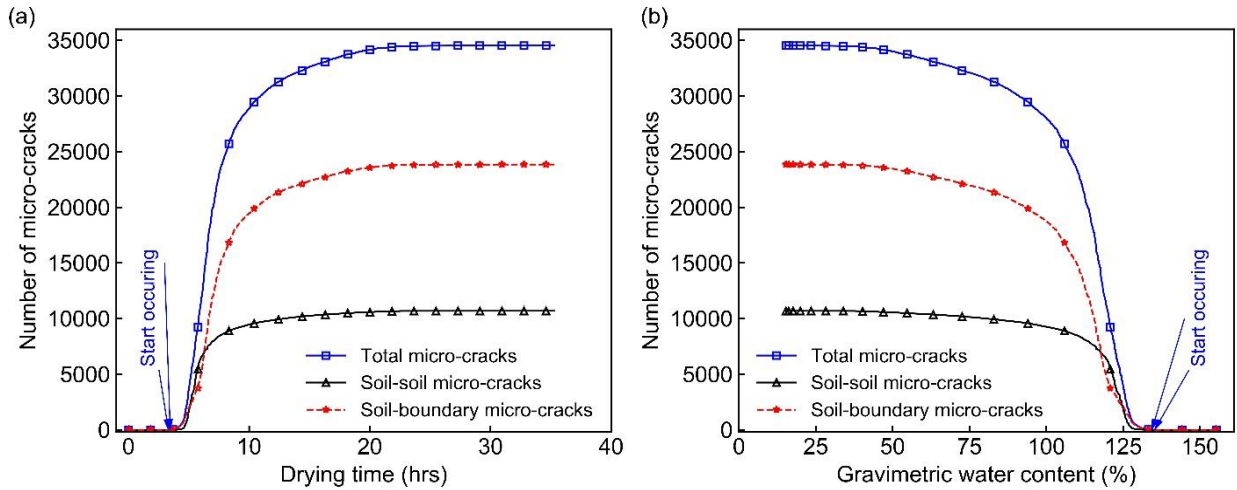

Fig. 22. Evolution of the number of micro-cracks during simulation of the circular sample: (a) number of micro-cracks against drying time; (b) number of micro-cracks against gravimetric water content.

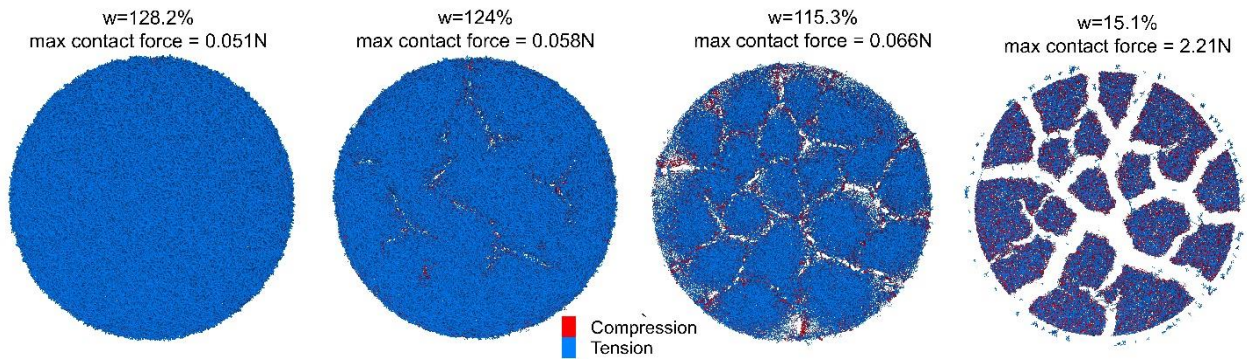

Fig. 23. Changes in the contact force network in the circular numerical sample during drying.

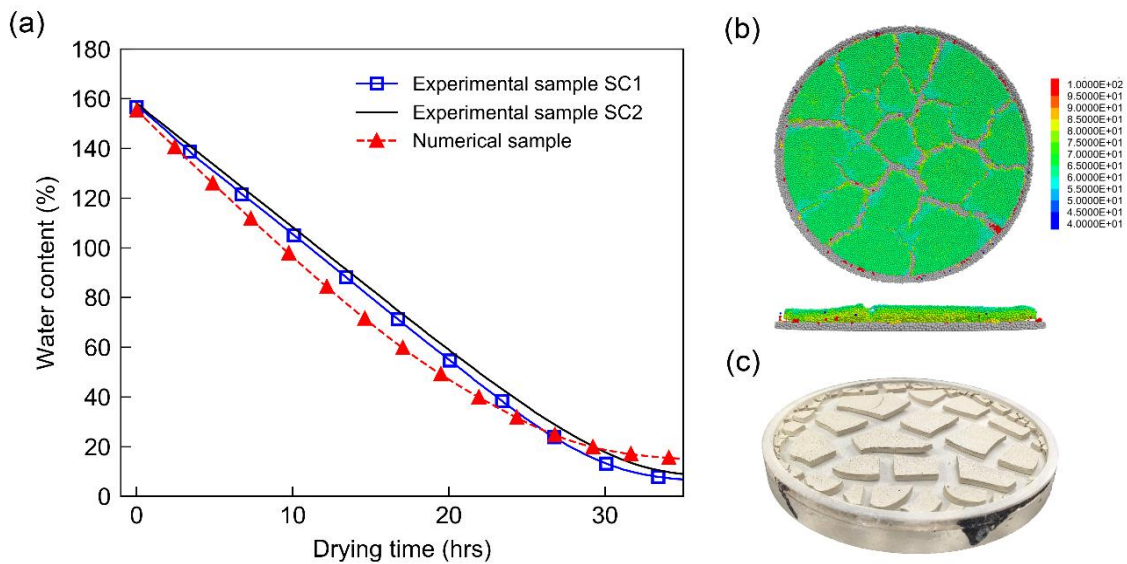

Fig. 24. (a) evolution of water content of the circular samples; (b) distribution of gravimetric water content (%)in the numerical sample (top and side views) at w = 71.6%; (c) sample SC1 after drying.

Fig. 24 shows the changes in water content of the numerical and experimental circular samples during drying, as well as its distribution in the numerical sample at the average water content of 71.6%, and experimental sample SC1 after drying. Changes in water content of the numerical sample are consistent with those of the experimental samples. The water content in all samples during drying exhibited a strong decline period, followed by a gradual decrease to a residual value (Fig. 24(a)). Fig. 24(b) shows that the water content at/on crack lips is slightly lower than other locations and the water content at locations near the sample surface is lower than that at deeper locations. Similar behaviour was reported in the literature when drying tests of clayey soil samples were conducted (Nahlawi and Kodikara 2006). Fig. 24(b) also shows the appearance of concave-up curling in the numerical sample. This type of soil curling is also observed in the experimental sample (Fig. 24(c)). The occurrence of concave-up curling may be attributed to the differential shrinkage along the height of the sample (Kodikara et al. 2004, Tran et al. 2019). This differential shrinkage occurs due to the nonuniform distribution of water content during the drying process of the sample. In this sample, as the evaporation process occurs on the sample surface, the particles on or near the sample surface drop their water content faster (as shown Fig. 24(b)), and thus these particles shrink faster. Further studies with the support of advanced techniques are required to capture the differential shrinkage or shrinkage anisotropy of the sample, while more information relating to soil curling behaviour can be found in literature (Kodikara et al. 2004, Tran et al. 2019).

## 4. Conclusions

This paper has presented a hybrid discrete-continuum numerical framework capable of capturing coupled hydro-mechanical behaviour for predicting soil desiccation. In this framework, an unsaturated soil layer is discretised into an assembly of DEM particles bonded together at their contacts, and its behaviour is governed by a cohesive model. This feature closely represents the mechanical behaviour of clayey soil at meso-level as well as ensuring that cracks due to mixed-mode failure are captured. To model the hydraulic behaviour of an unsaturated soil layer that can deform during drying, mathematical equations are derived based on the natural process of unsaturated flow at the micro-level, in which water flows along water layers adhering to solid particles. These equations also consider the deformation of the soil layer in the changes in the water content. A numerical algorithm is then proposed to link the hydraulic behaviour to the mechanical behaviour. The proposed framework is employed to predict the soil behaviour observed in the experimental samples during desiccation. The results show that the proposed framework can capture well both the hydraulic behaviour and the mechanical behaviour of the samples. It can also capture other typical behaviours of the soil desiccation cracking process, including the sequence of shrinkage and crack occurrence, as well as the occurrence of cracks due to mixed-mode failure, which have been not captured in previous studies using DEM. The results presented in this study suggest that the proposed framework can be used to provide further insights into the hydro-mechanical behaviour of soils in geotechnical applications.

**Acknowledgements**

Funding support from the Australian Research Council via projects DP160100775 (Ha H. Bui) and DP170103793 & DP190102779 (Ha H. Bui & Giang D. Nguyen) and FT200100884 (Ha H. Bui) is gratefully acknowledged. This research was undertaken with the assistance of resources and services from the National Computational Infrastructure (NCI), which is supported by the Australian Government.